\documentclass[12pt]{article}
\usepackage{epsfig,amssymb,amsmath,psfrag}

\catcode`\@=11
\DeclareMathAlphabet{\mathpzc}{OT1}{pzc}{m}{it}

\newcommand{\insertfig}[2]{\mbox{\epsfxsize=#1cm \epsfbox{#2.eps}}}

\font\cmss=cmss12 
\def\1{\hbox{{1}\kern-.25em\hbox{l}}}
\def\bfZ{\relax{\hbox{\cmss Z\kern-.4em Z}}}

\def \be  {\begin{equation}}
\def \ee  {\end{equation}}
\def \ba  {\begin{eqnarray}}
\def \ea  {\end{eqnarray}}
\def \baa {\begin{eqnarray*}}
\def \eaa {\end{eqnarray*}}
\def \bb  {\begin {thebibliography} }
\def \eb  {\end{thebibliography}}
\def \lab #1 {\label{#1}}

\newcommand\re[1]{(\ref{#1})}

\def \matrix #1 {\left(\begin{array}{cc} #1 \end{array}\right)}

\def \tr {\mathop{\rm tr}\nolimits}

\newcommand \vev [1] {\langle{#1}\rangle}

\newcommand{\as}{\ifmmode\alpha_{\rm s}\else{$\alpha_{\rm s}$}\fi}
\newcommand{\asbar}{\ifmmode\bar{\alpha}_{\rm s}\else{$\bar{\alpha}_{\rm s}$}\fi}

\newcommand{\bit}[1]{\mbox{\boldmath$#1$}}
\newcommand{\ft}[2]{{\textstyle\frac{#1}{#2}}}

\font\cmss=cmss12 
\def\inbar{\,\vrule height1.5ex width.4pt depth0pt}
\def\IC{\relax\hbox{$\inbar\kern-.3em{\rm C}$}}
\def\IZ{\relax{\hbox{\cmss Z\kern-.4em Z}}}
\def\IR{{\hbox{{\rm I}\kern-.2em\hbox{\rm R}}}}

\def\IP{{\hbox{{\rm I}\kern-.2em\hbox{\rm P}}}}
\def\II{\hbox{{1}\kern-.25em\hbox{l}}}

\def\numberbysection{\@addtoreset{equation}{section}
                     \def\theequation{\thesection.\arabic{equation}}}
\numberbysection

\newbox\lett\newdimen\lheight\newdimen\lwidth
\def\ontop#1#2{\setbox\lett=\hbox{#2}\lheight\ht\lett
\multiply\lheight by 12 \divide\lheight by 10\relax%
\lwidth\wd\lett \multiply\lwidth by 8 \divide\lwidth by 10\relax #2\kern-\lwidth%
\raise\lheight\hbox{{$\scriptstyle #1$}}\kern.1ex}

\def\XXint#1#2#3{{\setbox0=\hbox{$#1{#2#3}{\int}$}
     \vcenter{\hbox{$#2#3$}}\kern-.5\wd0}}

\begin{document}

\begin{titlepage}

\thispagestyle{empty}

\vskip2cm

\centerline{\large \bf Analytic Bethe Ansatz and Baxter equations}
\vspace{0.1cm}
\centerline{\large \bf for long-range $psl(2|2)$ spin chain}

\vspace{1cm}

\centerline{\sc A.V. Belitsky}

\vspace{10mm}

\centerline{\it Department of Physics, Arizona State University}
\centerline{\it Tempe, AZ 85287-1504, USA}

\vspace{1cm}

\centerline{\bf Abstract}

\vspace{5mm}

We study the largest particle-number-preserving sector of the dilatation operator in maximally
supersymmetric gauge theory. After exploring one-loop Bethe Ansatze for the underlying spin chain
with $psl(2|2)$ symmetry for simple root systems related to several Kac-Dynkin diagrams, we use
the analytic Bethe Anzats to construct eigenvalues of transfer matrices with finite-dimensional
atypical representations in the auxiliary space. We derive closed Baxter equations for eigenvalues
of nested Baxter operators. We extend these considerations for a non-distinguished root system
with FBBF grading to all orders of perturbation theory in 't Hooft coupling. We construct generating
functions for all transfer matrices with auxiliary space determined by Young supertableaux $(1^a)$
and $(s)$ and find determinant formulas for transfer matrices with auxiliary spaces corresponding to
skew Young supertableaux. The latter yields fusion relations for transfer matrices with auxiliary
space corresponding to representations labelled by square Young supertableaux. We derive asymptotic
Baxter equations which determine spectra of anomalous dimensions of composite Wilson operators in
noncompact $psl(2|2)$ subsector of $\mathcal{N} = 4$ super-Yang-Mills theory.

\end{titlepage}

\setcounter{footnote} 0

\newpage

\pagestyle{plain}
\setcounter{page} 1

\section{Introduction}

Low-dimensional integrable structures were long known to emerge in Quantum Chromodynamics --- the
theory of strong interaction. Evolution equations describing logarithmic modification of scattering
amplitudes in kinematical regimes corresponding to different physical phenomena were found to
possess hidden symmetries. These include high-energy and large-momentum transfer asymptotics of the
theory. The former refers to Regge behavior of cross sections in energy variable. In leading
logarithmic approximation it is governed by reggeon quantum mechanics, i.e., an interacting system
with conserved number of particles. Its Hamiltonian was identified with the one of a noncompact
Heisenberg magnet with $SL(2, \mathbb{C})$ symmetry group \cite{Lip93,FadKor95}. The regime of
scattering amplitudes with large momentum transfer is endowed with operator product expansion such
that evolution equations are equivalent to Callan-Symanzik equations. The latter is in turn a Ward
identity for dilatations which is one of the generators of the conformal group, the symmetry of
classical Lagrangian of the theory. The dilatation operator acts on the space spanned by Wilson
operators --- composite operators built from elementary fields of the theory and covariant
derivatives. At leading order of perturbation theory, the dilatation operator admits a pair-wise
form for a class of quasipartonic operators \cite{BukFroKurLip85}. In multicolor $N_c \to \infty$
limit, only nearest neighbor interactions survive with long-range effects being suppressed by $1/N_c$.
It was realized that the dilatation operator for aligned-helicity operators coincides with the
Hamiltonian of yet another integrable system --- spin chain with $SL(2, \mathbb{R})$ symmetry group
\cite{BraDerMan98,BraDerKorMan99,Bel99}. All four-dimensional gauge theories inherit integrability
of one-loop dilatation operator since they all share the same noncompact sector of operators with
covariant derivatives \cite{BelGorKor03,BelDerKorMan04}. Supersymmetry enhances the phenomenon to a
larger set of operators eventually encompassing all operators in maximally supersymmetric gauge
theory as was demonstrated in \cite{Lip98,MinZar03,BeiKriSta03,Bei03,BeiSta03,BelDerKorMan04}. The
super-spin chain Hamiltonian inherits the symmetry group of the classical gauge theory and its
spectrum can be found by means of the nested Bethe Ansatz \cite{Yan67}.

Beyond leading order of perturbation theory in a generic gauge theory many space-time charges
associated with classical symmetry generators cease to be conserved due to anomalies. This
may potentially lead to breaking of integrability in higher loops. The $\mathcal{N} = 4$
super-Yang-Mills theory, on the other hand, is superconformal to all orders and thus its classical
symmetry persists even when quantum effects are taken into account. However, starting from two-loop
order the dilatation operator becomes long-ranged, namely more then two partons can be simultaneously
involved in scattering. This invalidates standard procedures to derive Bethe Ansatz equations.
Evidence gathered from multiloop perturbative calculations hinted that integrability in maximally
supersymmetric gauge theory carries on to higher orders and that the spectrum of composite operators
is encoded in a long-range super-spin chain model \cite{BeiDipSta04,Sta04,BeiSta05}. Bethe Ansatz type
equations which depend on the 't Hooft coupling constant $g = g_{\rm\scriptscriptstyle YM} \sqrt{N_c}
/(2 \pi)$ and which generate known anomalous dimensions in lowest orders of perturbative expansion
were conjectured in Ref.\ \cite{BeiSta05}. Integrability of planar $\mathcal{N} = 4$ super Yang-Mills
theory then suggests that spectra of anomalous dimensions of composite operators can be computed
exactly at finite coupling $g$ and the superconformal nature of the theory implies that this provides
an exact solution to it. These Bethe Ansatz equations do not properly incorporate wrapping effect
when the range of the interaction becomes as long as the spin chain itself and thus the equations
are intrinsically asymptotic.

In this paper we will continue developing an alternative approach to long-range super-spin magnets based
on transfer matrices and Baxter equations \cite{Bax80} initiated in Refs.\ \cite{BelKorMul06,Bel06,Bel07}.
Transfer matrices is the main ingredient of this framework. They encode the full set of mutually
commuting conserved quantities with Hamiltonian being one of these. Transfer matrices are supertraces of
monodromy matrices in a representation of the symmetry algebra. The Baxter operators themselves are
certain transfer matrices with a special --- spectral parameter-dependent --- dimension of representations
in the auxiliary space. Thus, they are quantum generalization of supercharacters and we should expect a
transfer matrix (or its eigenvalues) to be a sum of terms, one per component of corresponding representation
in the auxiliary space. The absence of $R-$matrices yielding the putative long-range Bethe equations does
not prevent us from formulating transfer matrices of the model. To accomplish this goal we resort on the
analytic Bethe Ansatz \cite{Res83}, a techniques which bypasses the microscopic treatment and relies on
general properties of the macroscopic system like analyticity, unitarity and crossing symmetry.

Recently we have addressed Baxter equation for the closed $sl(2|1)$ subsector of the $\mathcal{N} = 4$
dilatation operator and have shown that it takes the form of a second order finite difference equation
\cite{Bel07} like for the $sl(2)$ long-range magnet \cite{Bel06}. In the present work, we will focus
on the maximal particle-number-preserving $psl(2|2)$ sector \cite{Bei04} of the maximally supersymmetric
gauge theory studied in a number of papers at one \cite{BeiKazSakZar05}, two \cite{Zwi06} and all
\cite{BeiSta05} orders, where a similar form of Baxter equations is anticipated \cite{BelKor05}.

The outline of the paper is as follows. In the next section we describe the $sl(2|2)$ subsector of
$\mathcal{N} = 4$ super-Yang-Mills theory which arises as a projection of the theory on the
light-cone and restricting the particle content to $\mathcal{N} = 2$ hypermultiplet. Then in section
\ref{ShortRangeSpinChain}, we analyze the short range spin chain describing the spectrum of one-loop
dilatation operator. We start with the distinguished basis and construct transfer matrices with
simplest representations in auxiliary space which suffice to formulate closed Baxter equations for
respective nested Baxter functions. Subsequently we perform a set of Weyl super-reflections with
respect to odd roots to obtain Bethe equations corresponding to Kac-Dynkin diagram with two isotropic
fermionic roots which allows for a natural generalization to all orders of perturbation theory. In section
\ref{TransferMatricesSymmetric}, we construct all transfer matrices with symmetric and antisymmetric
atypical representations in the auxiliary space and build a determinant representation for transfer
matrices labelled by a skew Young superdiagram. Subsequently we find finite difference equations for
Baxter polynomials. All previous considerations are generalized in section \ref{LongRangeSpinChain} to
all orders of perturbation within the framework of asymptotic analytic Bethe Ansatz. Several appendices
contain details of projection of $su(2,2|4)$ superconformal symmetry of maximally supersymmetric Yang-Mills
theory to its $psl(2|2)$ subsector, superspace realization of these algebras and Serre-Chevalley bases
for $psl(2|2)$ corresponding to distinguished and symmetric Kac-Dynkin diagrams. Finally we determine
numbers of roots of nested Baxter functions in terms of eigenvalues of Cartan generators.

\section{$sl(2|2)$ subsector of $\mathcal{N} = 4$ super-Yang-Mills}

To start with, let us specify the subsector of $\mathcal{N} = 4$ super-Yang-Mills theory which
will be the focus of our current study. Recall that the physical field content of the maximally
supersymmetric gauge theory can be accommodated in a single chiral light-cone $\mathcal{N} = 4$
superfield \cite{BriLinNil83,Man83,BelDerKorMan04},
\ba
\label{N4-superfield}
{\Phi} \left( x^\mu, \theta^A \right)
\!\!\!&=&\!\!\!
\partial_z^{-1} A (x^\mu) + \theta^A
\partial_z^{-1} \bar\lambda_A (x^\mu) + \frac{i}{2!} \theta^A \theta^B \bar \phi_{AB} (x^\mu)
\nonumber\\
&+&\!\!\! \frac{1}{3!} \varepsilon_{ABCD} \theta^A \theta^B \theta^C \lambda^D (x^\mu)
-
\frac{1}{4!} \varepsilon_{ABCD} \theta^A \theta^B \theta^C \theta^D \partial_z \bar{A} (x^\mu) \, .
\ea
depending on the bosonic four-vector $x^\mu = (z, x^+, \bit{x}_\perp)$ with its minus component
being chiral light-cone coordinate $z = x^- + \ft12 \bar\theta_A \theta^A$ and four Grassmann
variables $\theta^A$. Truncating this superfield in one of the superspace coordinates, say $\theta^1$,
one observes the particle content falls into two $\mathcal{N} = 2$ superfields \cite{BelDerKorMan04}
\be
\label{TruncationN4toN2}
{\Phi} (x^\mu, \theta^A) |_{\theta^1 = 0}
=
\Phi_{\scriptscriptstyle\rm G} (x^\mu, \theta^2, \theta^3)
+
\theta^4 \Phi_{\scriptscriptstyle\rm WZ} (x^\mu, \theta^2, \theta^3)
\, ,
\ee
with one encoding the $\mathcal{N} = 2$ gauge supermultiplet and another $\mathcal{N} = 2$ Wess-Zumino
hypermultiplet \cite{BelDerKorMan04},
\ba
\Phi_{\scriptscriptstyle\rm G} (x^\mu, \theta^2, \theta^3)
\!\!\!&=&\!\!\!
\partial_z^{-1} A (x^\mu) + \partial_z^{- 1} \theta^j \bar\lambda_j (x^\mu)
+
\frac{i}{2} \theta^j \theta^k \bar\phi_{jk} (x^\mu)
\, , \\
\Phi_{\scriptscriptstyle\rm WZ} (x^\mu, \theta^2, \theta^3)
\!\!\!&=&\!\!\!
\partial_z^{- 1} \bar\lambda_4 (x^\mu) + i \theta^j \bar\phi_{4j} (x^\mu)
+
\theta^2 \theta^3 \lambda^1 (x^\mu)
\, ,
\ea
where the summation runs over the remaining odd directions in superspace, i.e., $j, k = 2, 3$. It is
straightforward to identify the closed $sl(2|2)$ subsector \cite{Bei04} of the full theory as the
$\Psi_{\rm WZ}$ component of the $\mathcal{N} = 4$ light-cone superfield projected on the light-cone,
i.e., $x^\mu = (z, 0, \bit{0}_\perp)$. For further use, it is convenient to introduce new notations
for components of the hypermultiplet and odd direction of the $\mathcal{N} = 2$ superspace. Namely,
identifying $\theta$'s as $\vartheta^1 = \theta^2$, $\vartheta^2 = \theta^3$, the gaugino fields as
$\bar\lambda_4 = \bar\chi$, $\psi = \lambda^1$ and the $su(2)$ doublet of scalars as follows $\varphi_a
= (\bar{\phi}_{42}, \bar{\phi}_{43}) = (X, Z)$, we get the Wess-Zumino superfield in the form
\be
\label{N2WZsuperfield}
\Phi_{\scriptscriptstyle\rm WZ} (\mathcal{Z})
=
\partial_z^{- 1} \bar\chi (z) + i \vartheta^a \varphi_{a} (z)
+
\frac{1}{2} \varepsilon_{ab} \vartheta^a \vartheta^b \psi (z)
\, ,
\ee
depending on the light-cone superspace variable $\mathcal{Z} = (z, \vartheta^a)$. Truncating further
in either $\vartheta^1$ or $\vartheta^2$ variable, one gets the closed $sl(2|1)$ sector recently
discussed in Ref.\ \cite{Bel07}. Thus in the large-$N_c$ limit the $sl(2|2)$ sector of the dilatation
operator in the maximally supersymmetric Yang-Mills theory is spanned by single trace operators built
from the Wess-Zumino superfields,
\be
\label{LCoperator}
\mathbb{O} (\mathcal{Z}_1, \dots, \mathcal{Z}_L)
=
\tr
\left\{
\Phi_{\scriptscriptstyle\rm WZ} (\mathcal{Z}_1) \dots \Phi_{\scriptscriptstyle\rm WZ} (\mathcal{Z}_L)
\right\}
\, .
\ee

The light-cone superfield $\Phi_{\scriptscriptstyle\rm WZ} (\mathcal{Z})$ defines an
infinite-dimensional chiral representation of the $sl(2|2)$ algebra with generators realized
as differential operators acting on superspace coordinates as shown in Appendix
\ref{SuperspaceRealization}. The Wess-Zumino superfield possesses a vanishing conformal spin
and involves a nonlocal operator $\partial^{-1}_z \bar\chi (0)$ as an artefact of the light-cone
formalism. To overcome this complication we introduce a regularization\footnote{This
regularization affects the form of the generators \re{psl22} which receive additive addenda:
$\mathcal{L}^+ \to \mathcal{L}^+_\epsilon = \mathcal{L}^+ + 2 \epsilon z$, $\mathcal{L}^0 \to
\mathcal{L}^0_\epsilon = \mathcal{L}^0 + \epsilon$, $\bar{\mathcal{V}}^{a,+} \to
\bar{\mathcal{V}}^{a,+}_\epsilon = \bar{\mathcal{V}}^{a,+} + 2 \epsilon \vartheta^a$ and
$\mathcal{B} \to \mathcal{B}_\epsilon = \mathcal{B} + \ft12 \epsilon$.} by setting the conformal
dimension of $\Phi_{\scriptscriptstyle\rm WZ} (\mathcal{Z})$ to $\ell_{\scriptscriptstyle\rm WZ}
= \epsilon$ and taking the physical limit $\epsilon \to 0$ at the end. The representation space
$\mathbb{V}_\epsilon$ arises from the expansion of the superfield in Taylor series in even and
odd variables around $z = \vartheta^a = 0$ and is spanned by polynomials in $z$ and $\vartheta^a$,
\be
\mathbb{V}_\epsilon
=
{\rm span}
\{
1, z^{k + 1}, \vartheta^a z^k , \ft12 \varepsilon_{ab} \vartheta^a \vartheta^b z^k
| k \in \mathbb{N}
\}
\, .
\ee
The single-trace $L-$field light-cone operator \re{LCoperator} belongs to the $L-$fold product of these
spaces $\mathbb{V}_\epsilon^{\otimes L}$. Therefore, the eigenfunctions $\Psi_{\bit{\scriptstyle\omega}}$
of the spin chain are classified according to irreducible components entering this tensor product
parameterized by the eigenvalues $\bit{\omega} = [\ell, t, b, L]$ of generators of the $psl(2|2)$
Cartan subalgebra, $u_B (1)$ automorphism and the length of the operator (see Appendix
\ref{SerreChevalleyAppendix}). The local Wilson operators $\mathcal{O}_{\bit{\scriptstyle\omega}}$
corresponding to superconformal polynomials $\Psi_{\bit{\scriptstyle\omega}}$ can be projected out
from the light-cone operator \re{LCoperator} by means of the $sl(2|2)-$invariant scalar product,
\ba
\label{sl22ScalarProduct}
\mathcal{O}_{\bit{\scriptstyle\omega}}
\!\!\!&=&\!\!\!
\vev{\Psi_{\bit{\scriptstyle \omega}} (\mathcal{Z}_1, \dots, \mathcal{Z}_L)|
\mathbb{O} (\mathcal{Z}_1, \dots, \mathcal{Z}_L)}
\nonumber\\
&\equiv&\!\!\!
\int
\prod_{k = 1}^L [\mathcal{D} \mathcal{Z}_k ]_\epsilon \,
\overline{\Psi_{\bit{\scriptstyle \omega}}
(\mathcal{Z}_1, \dots, \mathcal{Z}_L)} \, \mathbb{O} (\mathcal{Z}_1, \dots, \mathcal{Z}_L)
\, ,
\ea
with the integration measure \cite{BelDerKorMan05}
\be
\label{sl22measure}
\int [\mathcal{D} \mathcal{Z} ]_\epsilon
=
\frac1{\Gamma(2\epsilon + 1)}
\int_{|z|\le 1} \frac{d^2 z}{\pi} \int \prod_{a = 1}^2 \left(d \bar\vartheta_a d\vartheta^a \right) \,
(1 - \bar{z} z + \bar\vartheta_a \vartheta^a)^{2 \epsilon}
\, .
\ee
One can easily convince oneself that the polynomials $z^k \vartheta^a \dots$, forming the representation
space $\mathbb{V}_\epsilon$, are orthogonal with respect to the $sl(2|2)-$scalar product, i.e., $\langle z^k
\vartheta^a \dots | z^n \vartheta^b \dots \rangle \sim \delta_{kn} \delta_a^b \dots$. The nonlocal
operator $\partial_z^{-1} \bar\chi (0)$ associated with the lowest component in the Taylor expansion
of the superfield $\Phi_{\scriptscriptstyle\rm WZ} (Z)$ defines an invariant one-dimensional subspace
$\mathbb{V}_{\rm nonloc} = \{ 1 \}$. Local Wilson operators belong to the quotient $\mathbb{V}_{\rm loc}
= \mathbb{V}_\epsilon / \mathbb{V}_{\rm nonloc}$. With the chosen normalization of the measure
\re{sl22measure}, one finds that $\langle z^k | z^n \rangle \sim \delta_{kn} / \Gamma (2 \epsilon + k)$
such that the vector belonging to $\mathbb{V}_{\rm nonloc}$ possesses zero norm and is orthogonal to
all states
\be
\{ z^{k + 1}, \vartheta^1 z^k, \vartheta^2 z^k, \vartheta^1 \vartheta^2 z^k | k \in \mathbb{N} \}
\leftrightarrow
\{
( \mathcal{D}^+ )^k \bar\chi ,
( \mathcal{D}^+ )^k X ,
( \mathcal{D}^+ )^k Z ,
( \mathcal{D}^+ )^k \psi
| k \in \mathbb{N} \}
\, ,
\ee
belonging to $\mathbb{V}_{\rm loc}$. These will be identified with excitations propagating on the super-spin
chain which we will turn to next.

\section{Short-range spin chain}
\label{ShortRangeSpinChain}

\begin{figure}[t]
\begin{center}
\mbox{
\begin{picture}(0,40)(200,0)
\put(0,0){\insertfig{14}{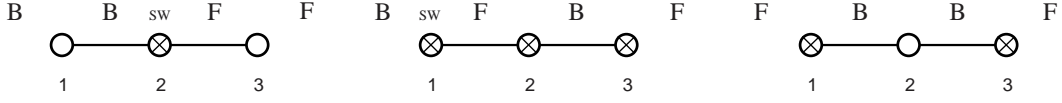}}
\end{picture}
}
\end{center}
\caption{\label{dynkin} A subset of Kac-Dynkin diagrams for $psl(2|2)$. The super-Weyl
reflection of the nodes denoted by the label ${\rm\scriptstyle SW}$ generates the diagram
standing to its right.}
\end{figure}

The eigenvalue problem for the spectrum of one-loop anomalous dimensions of superconformal
operators \re{LCoperator} can be reformulated in terms of a short-range $psl(2|2)$ quantum
super-spin chain with the one-loop dilatation operator being identified with the
nearest-neighbor Hamiltonian of the magnet. The spin chain is integrable and it can be
diagonalized be means of the nested Bethe Ansatz \cite{Yan67}. It was observed some time
ago \cite{Kul86,ResWie87} that for a spin chain model based on a given symmetry (super-)algebra,
the nested Bethe Ansatz equations are determined by simple root systems of the algebra, generally,
\be
\label{NestedABAgeneric}
(- 1)^{ A_{pp}/2} \left(
\frac{u^{(p)}_{0,k} - \ft{i}{2} w_p}{u^{(p)}_{0,k} + \ft{i}{2} w_p}
\right)^L
=
\prod_{q = 1}^{N_r} \prod_{j = 1}^{n_q}
\frac{
u^{(p)}_{0,k} - u^{(q)}_{0,j} + \ft{i}{2} A_{pq}
}{
u^{(p)}_{0,k} - u^{(q)}_{0,j} - \ft{i}{2} A_{pq}
}
\, ,
\ee
where\footnote{See Appendix \ref{SerreChevalleyAppendix} for details.} $A_{pq} = (\bit{\alpha}_p |
\bit{\alpha}_q)$ is the Cartan matrix and $w_p$ are the Kac-Dynkin labels, $w_p = (\bit{\alpha}_p|
\bit{\mu})$ determined by a weight vector $\bit{\mu}$ of a representation of the algebra acting on
the spin chain sites. Here $N_r = {\rm rank} (G)$ is the rank of the algebra $G$. For projective
algebras the upper limit in the product is $N_r = {\rm rank} (G) + 1$, which is $3$ for our $psl(2|2)$
sector. As it becomes obvious from the above equation, there exists several sets of equivalent Bethe
Ansatz equations reflecting the fact that there are several choices of simple root systems
$\{\bit{\alpha}_p |p = 1, {\dots}, N_r \}$ for a superalgebra, see Fig.\ \ref{dynkin}. These simple
root systems are related by reflections with respect to odd simple roots $\bit{\alpha}$ with vanishing
bilinear form $(\bit{\alpha}|\bit{\alpha}) = 0$ which form the Weyl supergroup. For Bethe equations
this corresponds to a particle-hole transformations.

To diagonalize the short-range $psl(2|2)$ magnet one can use the nested Algebraic Bethe Ansatz
\cite{Kul86,KulSkl82} following Refs.\ \cite{EssKorSch92,EssKor92,Sch93} and construct transfer
matrices by a fusion procedure \cite{KulResSkl81}. However, the lack of a systematic procedure to
construct long-range integrable spin chains corresponding to gauge theories, binds one has to resort
to techniques which bypass the microscopic treatment and rely on general properties of macroscopic
systems. The method of analytic Bethe Ansatz, which is a generalization of the inverse scattering
method, was developed to determine the spectrum of transfer matrices for closed chains \cite{Res83}
and serves the purpose. In this approach, one uses general properties such as analyticity, unitarity,
crossing symmetry, etc., to completely determine eigenvalues of transfer matrices.

\subsection{Transfer matrices and Baxter equations in distinguished basis}
\label{TransferBaxterDistinguished}

Let us start with the distinguished Kac-Dynkin (left-most) diagram in Fig.\ \ref{dynkin}
with BBFF grading encoded in the Cartan matrix
\be
\label{CartanDistinguished}
A =
\left(
\begin{array}{rrr}
2  & -1 &  0 \\
-1 &  0 &  1 \\
0  &  1 & -2
\end{array}
\right)
\, .
\ee
The nested Bethe equations read, according to Eq.\ \re{NestedABAgeneric},
\ba
\label{DistinguishedBAElo}
- \left( \frac{\tilde{u}^{(1)}_{0,k} - i}{\tilde{u}^{(1)}_{0,k} + i} \right)^L
\!\!\!&=&\!\!\!
\frac{
\widetilde{Q}^{(1)}_0 \left( \tilde{u}^{(1)}_{0,k} + i \right)
}{
\widetilde{Q}^{(1)}_0 \left( \tilde{u}^{(1)}_{0,k} - i \right)
}
\frac{
\widetilde{Q}^{(2)}_0 \left(\tilde{u}^{(1)}_{0,k} - \ft{i}{2} \right)
}{
\widetilde{Q}^{(2)}_0 \left(\tilde{u}^{(1)}_{0,k} + \ft{i}{2} \right)
}
\, , \nonumber\\
\left( \frac{\tilde{u}^{(2)}_{0,k} + \ft{i}{2}}{\tilde{u}^{(2)}_{0,k} - \ft{i}{2}} \right)^L
\!\!\!&=&\!\!\!
\frac{
\widetilde{Q}^{(1)}_0 \left(\tilde{u}^{(2)}_{0,k} - \ft{i}{2} \right)
}{
\widetilde{Q}^{(1)}_0 \left(\tilde{u}^{(2)}_{0,k} + \ft{i}{2} \right)
}
\frac{
\widetilde{Q}^{(3)}_0 \left(\tilde{u}^{(2)}_{0,k} + \ft{i}{2} \right)
}{
\widetilde{Q}^{(3)}_0 \left(\tilde{u}^{(2)}_{0,k} - \ft{i}{2} \right)
}
\, , \\
- 1 \!\!\!&=&\!\!\!
\frac{
\widetilde{Q}^{(3)}_0 \left( \tilde{u}^{(3)}_{0,k} - i \right)
}{
\widetilde{Q}^{(3)}_0 \left( \tilde{u}^{(3)}_{0,k} + i \right)
}
\frac{
\widetilde{Q}^{(2)}_0 \left(\tilde{u}^{(3)}_{0,k} + \ft{i}{2} \right)
}{
\widetilde{Q}^{(2)}_0 \left(\tilde{u}^{(3)}_{0,k} - \ft{i}{2} \right)
}
\, . \nonumber
\ea
They are written in terms of the Baxter polynomials
\be
\widetilde{Q}^{(p)}_0 (u) = \prod_{k = 1}^{\tilde{n}_p} \left( u - \tilde{u}^{(p)}_{0,k} \right)
\, ,
\ee
parameterized by three sets of Bethe roots $\tilde{u}^{(p)}_{0,k}$. Here and below in this section
all symbols carry a subscript $0$ indicating zeroth order of perturbation theory for the corresponding
quantities in gauge theory. These equations yield one-loop anomalous dimensions for Wilson operators
in $psl(2|2)$ sector of $\mathcal{N} = 4$ super-Yang-Mills theory.

In this section we briefly discuss the construction of eigenvalues for transfer matrices in
the distinguished basis echoing considerations of Ref.\ \cite{Tsu97} adopted to $psl(2|2)$
algebra. Transfer matrices are supertraces of monodromy matrices with certain representation of
symmetry algebra in the auxiliary space. Here we will present only the ones with low-dimensional
representations in the auxiliary space which are sufficient to derive closed Baxter equations for
nested Baxter polynomials. A full-fledge considerations will be done below for a simple root
system corresponding to the symmetric Kac-Dynkin diagram in FBBF grading with two isotropic fermionic
roots, see (right-most graph) Fig.\ \ref{dynkin}. Being supertraces over a representation of the
algebra, the eigenvalue formulas are expected to be given by a sum of terms, one for each component
of the representation. This idea is at the crux of the approach suggested for spin chain based on
classical Lie algebras in Ref.\ \cite{BazRes89,KunSuz95} and generalized for superalgebras in Ref.\
\cite{Tsu97}.

We derive transfer matrices with auxiliary space labelled by a particular Young supertableau.
The Young superdiagrams are different from classical ones in that there is no limitation on
the number of rows \cite{BalBar81}. Thus, we introduce elementary Young supertableaux depending
on a spectral parameter $u$
\ba
\label{DistinguishedBox}
\unitlength0.4cm
\begin{picture}(2.3,1.2)
\linethickness{0.4mm}
\put(1,0){\framebox(1,1){$1$}}
\end{picture}
{}_{u}
\!\!\!&=&\!\!\!
\left( u + \ft{i}{2} \right)^L
\frac{
\widetilde{Q}^{(1)}_0 \left( u + \ft{i}{2} \right)}{\widetilde{Q}^{(1)}_0 \left( u - \ft{i}{2} \right)}
\, , \\
\unitlength0.4cm
\begin{picture}(2.3,1)
\linethickness{0.4mm}
\put(1,0){\framebox(1,1){$2$}}
\end{picture}
{}_{u}
\!\!\!&=&\!\!\!
\left( u - \ft{3 i}{2} \right)^L
\frac{\widetilde{Q}^{(1)}_0 \left( u - \ft{3 i}{2} \right)}{\widetilde{Q}^{(1)}_0 \left( u - \ft{i}{2} \right)}
\frac{\widetilde{Q}^{(2)}_0 (u)}{\widetilde{Q}^{(2)}_0 (u - i)}
\, , \nonumber\\
\unitlength0.4cm
\begin{picture}(2.3,1)
\linethickness{0.4mm}
\put(1,0){\framebox(1,1){$3$}}
\end{picture}
{}_{u}
\!\!\!&=&\!\!\!
\left( u - \ft{i}{2} \right)^L
\frac{\widetilde{Q}^{(3)}_0 \left( u - \ft{3 i}{2} \right)}{\widetilde{Q}^{(3)}_0 \left( u - \ft{i}{2} \right)}
\frac{\widetilde{Q}^{(2)}_0 (u)}{\widetilde{Q}^{(2)}_0 (u - i)}
\, , \nonumber\\
\unitlength0.4cm
\begin{picture}(2.3,1)
\linethickness{0.4mm}
\put(1,0){\framebox(1,1){$4$}}
\end{picture}
{}_{u}
\!\!\!&=&\!\!\!
\left( u - \ft{i}{2} \right)^L
\frac{\widetilde{Q}^{(3)}_0 \left( u + \ft{i}{2} \right)}{\widetilde{Q}^{(3)}_0 \left( u - \ft{i}{2} \right)}
\, , \nonumber
\ea
parameterized in terms of three Baxter polynomials $\widetilde{Q}^{(k)}_0$ entering the nested Bethe
Ansatz equations \re{DistinguishedBAElo}. Each box is labelled by an index with grading $\bar{1} =
\bar{2} = 0$ and $\bar{3} = \bar{4} = 1$ in accord with the distinguished Kac-Dynkin diagram in Fig.\
\ref{dynkin}.

We introduce notations for eigenvalues of transfer matrices
\be
t_{0,(1^a)} (u) = t_{0,[a]} (u)
\, , \qquad
t_{0,(s)} (u) = t_0^{\{s\}} (u)
\, ,
\ee
with totally antisymmetric $(1^a)$ and symmetric $(s)$ atypical representations in the auxiliary space. For
the lowest dimensional representations, the eigenvalues of transfer matrices can we written in terms of the
elementary boxes as follows\footnote{Notice that the transfer matrix $t_{0,[1]} (u)$ with defining fundamental
representation $(1)$ in the auxiliary space is given by the supertrace ${\rm str} [\mathbb{L}_L (u) {\dots}
\mathbb{L}_1 (u)]$ of the product of Lax operators $\mathbb{L} (u)$ \cite{EssKorSch92,EssKor92,BelDerKorMan04}
and its eigenvalues in nested Bethe Ansatz given by Eq.\ \re{Fundamental1Tranfer}.}
\ba
\label{Fundamental1Tranfer}
t_{0,[1]} (u)
\!\!\!&=&\!\!\!
t_0^{\{1\}} (u)
=
\unitlength0.4cm
\begin{picture}(2.3,1)
\linethickness{0.4mm}
\put(1,-0.2){\framebox(1,1){$1$}}
\end{picture}
{}_{u}
+\!\!
\unitlength0.4cm
\begin{picture}(2.3,1)
\linethickness{0.4mm}
\put(1,-0.2){\framebox(1,1){$2$}}
\end{picture}
{}_{u}
-\!\!
\unitlength0.4cm
\begin{picture}(2.3,1)
\linethickness{0.4mm}
\put(1,-0.2){\framebox(1,1){$3$}}
\end{picture}
{}_{u}
-\!\!
\unitlength0.4cm
\begin{picture}(2.3,1)
\linethickness{0.4mm}
\put(1,-0.2){\framebox(1,1){$4$}}
\end{picture}
{}_{u}
\, , \\ [3mm]
t_{0,[2]} (u)
\!\!\!&=&\!\!\!\!\!\!
\unitlength0.4cm
\begin{picture}(2.3,1)
\linethickness{0.4mm}
\put(1,0.5){\framebox(1,1){$1$}}
\put(1,-0.6){\framebox(1,1){$2$}}
\end{picture}
\!\!
{\begin{array}{c} \scriptstyle u - \frac{i}{2} \vspace{-0.5mm} \\ \scriptstyle u + \frac{i}{2} \end{array}}
\!-\!\!\!\!
\unitlength0.4cm
\begin{picture}(2.3,1)
\linethickness{0.4mm}
\put(1,0.5){\framebox(1,1){$1$}}
\put(1,-0.6){\framebox(1,1){$3$}}
\end{picture}
\!\!
{\begin{array}{c} \scriptstyle u - \frac{i}{2} \vspace{-0.5mm} \\ \scriptstyle u + \frac{i}{2} \end{array}}
\!-\!\!\!\!
\unitlength0.4cm
\begin{picture}(2.3,1)
\linethickness{0.4mm}
\put(1,0.5){\framebox(1,1){$1$}}
\put(1,-0.6){\framebox(1,1){$4$}}
\end{picture}
\!\!
{\begin{array}{c} \scriptstyle u - \frac{i}{2} \vspace{-0.5mm} \\ \scriptstyle u + \frac{i}{2} \end{array}}
\!-\!\!\!\!
\unitlength0.4cm
\begin{picture}(2.3,1)
\linethickness{0.4mm}
\put(1,0.5){\framebox(1,1){$2$}}
\put(1,-0.6){\framebox(1,1){$3$}}
\end{picture}
\!\!
{\begin{array}{c} \scriptstyle u - \frac{i}{2} \vspace{-0.5mm} \\ \scriptstyle u + \frac{i}{2} \end{array}}
\!-\!\!\!\!
\unitlength0.4cm
\begin{picture}(2.3,1)
\linethickness{0.4mm}
\put(1,0.5){\framebox(1,1){$2$}}
\put(1,-0.6){\framebox(1,1){$4$}}
\end{picture}
\!\!
{\begin{array}{c} \scriptstyle u - \frac{i}{2} \vspace{-0.5mm} \\ \scriptstyle u + \frac{i}{2} \end{array}}
\!+\!\!\!\!
\unitlength0.4cm
\begin{picture}(2.3,1)
\linethickness{0.4mm}
\put(1,0.5){\framebox(1,1){$3$}}
\put(1,-0.6){\framebox(1,1){$4$}}
\end{picture}
\!\!
{\begin{array}{c} \scriptstyle u - \frac{i}{2} \vspace{-0.5mm} \\ \scriptstyle u + \frac{i}{2} \end{array}}
\!+\!\!\!\!
\unitlength0.4cm
\begin{picture}(2.3,1)
\linethickness{0.4mm}
\put(1,0.5){\framebox(1,1){$3$}}
\put(1,-0.6){\framebox(1,1){$3$}}
\end{picture}
\!\!
{\begin{array}{c} \scriptstyle u - \frac{i}{2} \vspace{-0.5mm} \\ \scriptstyle u + \frac{i}{2} \end{array}}
\!+\!\!\!\!
\unitlength0.4cm
\begin{picture}(2.3,1)
\linethickness{0.4mm}
\put(1,0.5){\framebox(1,1){$4$}}
\put(1,-0.6){\framebox(1,1){$4$}}
\end{picture}
\!\!
{\begin{array}{c} \scriptstyle u - \frac{i}{2} \vspace{-0.5mm} \\ \scriptstyle u + \frac{i}{2} \end{array}}
\, , \nonumber\\ [3mm]
t_0^{\{2\}} (u)
\!\!\!&=&\!\!\!\!\!\!
\unitlength0.4cm
\begin{picture}(4,1)
\linethickness{0.4mm}
\put(1,0){\framebox(1,1){$1$}}
\put(2.1,0){\framebox(1,1){$2$}}
\put(0.4,-1.3){$\scriptstyle u + \frac{i}{2}$}
\put(2.2,-1.3){$\scriptstyle u - \frac{i}{2}$}
\end{picture}
\!\!-\!\!
\unitlength0.4cm
\begin{picture}(4,1)
\linethickness{0.4mm}
\put(1,0){\framebox(1,1){$1$}}
\put(2.1,0){\framebox(1,1){$3$}}
\put(0.4,-1.3){$\scriptstyle u + \frac{i}{2}$}
\put(2.2,-1.3){$\scriptstyle u - \frac{i}{2}$}
\end{picture}
\!\!-\!\!
\unitlength0.4cm
\begin{picture}(4,1)
\linethickness{0.4mm}
\put(1,0){\framebox(1,1){$1$}}
\put(2.1,0){\framebox(1,1){$4$}}
\put(0.4,-1.3){$\scriptstyle u + \frac{i}{2}$}
\put(2.2,-1.3){$\scriptstyle u - \frac{i}{2}$}
\end{picture}
\!\!-\!\!
\begin{picture}(4,1)
\linethickness{0.4mm}
\put(1,0){\framebox(1,1){$2$}}
\put(2.1,0){\framebox(1,1){$3$}}
\put(0.4,-1.3){$\scriptstyle u + \frac{i}{2}$}
\put(2.2,-1.3){$\scriptstyle u - \frac{i}{2}$}
\end{picture}
\!\!-\!\!
\unitlength0.4cm
\begin{picture}(4,1)
\linethickness{0.4mm}
\put(1,0){\framebox(1,1){$2$}}
\put(2.1,0){\framebox(1,1){$4$}}
\put(0.4,-1.3){$\scriptstyle u + \frac{i}{2}$}
\put(2.2,-1.3){$\scriptstyle u - \frac{i}{2}$}
\end{picture}
\!\!+\!\!
\begin{picture}(4,1)
\linethickness{0.4mm}
\put(1,0){\framebox(1,1){$3$}}
\put(2.1,0){\framebox(1,1){$4$}}
\put(0.4,-1.3){$\scriptstyle u + \frac{i}{2}$}
\put(2.2,-1.3){$\scriptstyle u - \frac{i}{2}$}
\end{picture}
\!\!+ \!\!
\unitlength0.4cm
\begin{picture}(4,1)
\linethickness{0.4mm}
\put(1,0){\framebox(1,1){$1$}}
\put(2.1,0){\framebox(1,1){$1$}}
\put(0.4,-1.3){$\scriptstyle u + \frac{i}{2}$}
\put(2.2,-1.3){$\scriptstyle u - \frac{i}{2}$}
\end{picture}
\!\!+ \!\!
\unitlength0.4cm
\begin{picture}(4,1)
\linethickness{0.4mm}
\put(1,0){\framebox(1,1){$2$}}
\put(2.1,0){\framebox(1,1){$2$}}
\put(0.4,-1.3){$\scriptstyle u + \frac{i}{2}$}
\put(2.2,-1.3){$\scriptstyle u - \frac{i}{2}$}
\end{picture}
\, . \nonumber\\
\nonumber
\ea
The right-hand sides of these expressions are free from pole at positions of Bethe roots as can be
easily proved making use of the nested Bethe Ansatz equations \re{DistinguishedBAElo}. When written
explicitly in terms of Baxter polynomials, the conjugate transfer matrices, i.e., with antichiral
representations in the auxiliary space, can be obtained from these by merely dressing transfer matrices
with a bar and changing the signs in front of imaginary units. The generating function of all transfer
matrices will be given below in Sect.\ \ref{TransferMatricesSymmetric} though for the symmmetric
Kac-Dynkin diagram.

Using these transfer matrices and their conjugate one may derive closed equations obeyed by
the Baxter polynomials. First, solving the transfer matrix $t_{0,[1]} (u)$ with respect to
$\widetilde{Q}^{(1)}_0$ and substituting it into $\bar{t}_1 (u)$, one finds the Baxter
equation for $\widetilde{Q}^{(2)}_0$
\be
\label{DistBasisQ2}
t_{0,[1]} \left( u + \ft{i}{2} \right) \widetilde{Q}^{(2)}_0 \left( u - \ft{i}{2} \right)
-
\bar{t}_{0,[1]} \left( u - \ft{i}{2} \right) \widetilde{Q}^{(2)}_0 \left( u + \ft{i}{2} \right)
= 0
\, .
\ee
It is a first order finite-difference equation. Analogously, eliminating the polynomial
$\widetilde{Q}^{(3)}_0$ from the transfer matrix $t_{0,[1]}$ by substituting it twice into
$t_{0,[2]}$ and shifting its argument in middle of the way, one finds
\ba
\label{DistBasisQ2Q1}
\left[ t_{0,[2]} (u) - t_{0,[1]} \left( u + \ft{i}{2} \right) t_{0,[1]} \left( u - \ft{i}{2} \right) \right]
\widetilde{Q}^{(1)}_0 (u)
\!\!\!&+&\!\!\!
(u + i)^L t_{0,[1]} \left( u + \ft{i}{2} \right)
\widetilde{Q}^{(1)}_0 (u - i)
\\
&+&\!\!\!
(u - i)^L t_{0,[1]} \left( u - \ft{i}{2} \right)
\frac{
\widetilde{Q}^{(2)}_0 \left( u + \ft{i}{2} \right)
}{
\widetilde{Q}^{(2)}_0 \left( u - \ft{i}{2} \right)
}
\widetilde{Q}^{(1)}_0 (u - i)
= 0
\, . \nonumber
\ea
From this one can find a closed equation for the Baxter polynomial $\widetilde{Q}^{(1)}_0$ by merely
eliminating the ratio of the polynomials $\widetilde{Q}^{(2)}_0$ making use of Eq.\ \re{DistBasisQ2}.
An alternative, symmetric form of the Baxter equation can be found by deriving first an equation
analogous to \re{DistBasisQ2Q1} but with conjugate transfer matrices and eliminating the ratio of
$\widetilde{Q}^{(2)}_0$'s from them. This yields
\ba
\label{Baxterq1}
\left[ t_{0,[2]} (u) - t_{0,[1]} \left( u + \ft{i}{2} \right) t_{0,[1]} \left( u - \ft{i}{2} \right) \right]
\!\!\!\!&&\!\!\!\!\!
\left[ \bar{t}_{0,[2]} (u) - \bar{t}_{0,[1]} \left( u + \ft{i}{2} \right) \bar{t}_{0,[1]} \left( u - \ft{i}{2} \right) \right]
\widetilde{Q}^{(1)}_0 (u)
\\
+ \,
(u + i)^L
t_{0,[1]} \left( u - \ft{i}{2} \right)
\!\!\!\!&&\!\!\!\!\!
\left[ \bar{t}_{0,[2]} (u) - \bar{t}_{0,[1]} \left( u + \ft{i}{2} \right) \bar{t}_{0,[1]} \left( u - \ft{i}{2} \right) \right]
\widetilde{Q}^{(1)}_0 (u + i)
\nonumber\\
+ \,
(u - i)^L
\bar{t}_{0,[1]} \left( u + \ft{i}{2} \right)
\!\!\!\!&&\!\!\!\!\!
\left[ t_{0,[2]} (u) - t_{0,[1]} \left( u + \ft{i}{2} \right) t_{0,[1]} \left( u - \ft{i}{2} \right) \right]
\widetilde{Q}^{(1)}_0 (u - i)
= 0
\, . \nonumber
\ea
Similarly, solving for $\widetilde{Q}^{(1)}_0$ from $t_{0,[1]}$ and eliminating it from $t_{0,[2]}$ we
find an equation
\be
t_{0,[2]} (u) \widetilde{Q}^{(3)}_0 (u)
+
u^L t_{0,[1]} \left( u - \ft{i}{2} \right) \widetilde{Q}^{(3)}_0 (u + i)
+
u^L t_{0,[1]} \left( u - \ft{i}{2} \right)
\frac{
\widetilde{Q}^{(2)}_0 \left( u + \ft{i}{2} \right)
}{
\widetilde{Q}^{(2)}_0 \left( u - \ft{i}{2} \right)
}
\widetilde{Q}^{(3)}_0 (u - i)
= 0
\, ,
\ee
which gives a closed equation for $\widetilde{Q}^{(3)}_0$ upon the elimination of the polynomials
$\widetilde{Q}^{(2)}_0$,
\be
\label{BaxterQ3tildeDistinguished}
t_{0,[2]} (u) \bar{t}_{0,[2]} (u) \widetilde{Q}^{(3)}_0 (u)
+
u^L \bar{t}_{0,[2]} (u) t_{0,[1]} \left( u - \ft{i}{2} \right) \widetilde{Q}^{(3)}_0 (u + i)
+
u^L t_{0,[2]} (u) \bar{t}_{0,[1]} \left( u + \ft{i}{2} \right) \widetilde{Q}^{(3)}_0 (u - i)
= 0
\, .
\ee

So far we have derived Baxter equations in terms of transfer matrices with antisymmetric representation
in the auxiliary space. The same considerations can be performed with symmetric transfer matrices.
From symmetric matrices we find an analogue to Eq.\ \re{DistBasisQ2Q1} for $\widetilde{Q}^{(1)}_0$,
\be
t_0^{\{2\}} (u) \widetilde{Q}^{(1)}_0 (u)
-
(u + i)^L t_{0,[1]} \left( u - \ft{i}{2} \right) \widetilde{Q}^{(1)}_0 (u + i)
-
(u - i)^L t_{0,[1]} \left( u - \ft{i}{2} \right)
\frac{
\widetilde{Q}^{(2)}_0 \left( u + \ft{i}{2} \right)
}{
\widetilde{Q}^{(2)}_0 \left( u - \ft{i}{2} \right)
}
\widetilde{Q}^{(1)}_0 (u - i)
= 0
\, .
\ee
Eliminating $\widetilde{Q}^{(2)}_0$, we get yet another Baxter equation
\ba
t_0^{\{2\}} (u) \bar{t}_0^{\{2\}} (u) \widetilde{Q}^{(1)}_0 (u)
\!\!\!&-&\!\!\!
(u + i)^L \bar{t}_0^{\{2\}} (u) t_{0,[1]} \left( u - \ft{i}{2} \right) \widetilde{Q}^{(1)}_0 (u + i)
\\
&-&\!\!\!
(u - i)^L t_0^{\{2\}} (u) \bar{t}_{0,[1]} \left( u + \ft{i}{2} \right) \widetilde{Q}^{(1)}_0 (u - i)
= 0
\, , \nonumber
\ea
cf.\ Eq.\ \re{Baxterq1}. Finally, solving the system of transfer matrices $t_{0,[1]}$ and $t_0^{\{2\}}$
for $\widetilde{Q}^{(1)}_0$, we get
\ba
\left[ t_0^{\{2\}} (u) - t_{0,[1]} \left( u + \ft{i}{2} \right) t_{0,[1]} \left( u - \ft{i}{2} \right) \right]
\widetilde{Q}^{(3)}_0 (u)
\!\!\!&-&\!\!\!
u^L t_{0,[1]} \left( u - \ft{i}{2} \right)
\widetilde{Q}^{(3)}_0 (u + i)
\\
&-&\!\!\!
u^L t_{0,[1]} \left( u - \ft{i}{2} \right)
\frac{
\widetilde{Q}^{(2)}_0 \left( u + \ft{i}{2} \right)
}{
\widetilde{Q}^{(2)}_0 \left( u - \ft{i}{2} \right)
}
\widetilde{Q}^{(3)}_0 (u - i)
= 0
\, , \nonumber
\ea
which being solved for $\widetilde{Q}^{(2)}_0$ together with its conjugate gives
\ba
\left[ t_0^{\{2\}} (u) - t_{0,[1]} \left( u + \ft{i}{2} \right) t_{0,[1]} \left( u - \ft{i}{2} \right) \right]
\!\!\!\!&&\!\!\!\!\!
\left[ \bar{t}_0^{\{2\}} (u) - \bar{t}_{0,[1]} \left( u + \ft{i}{2} \right) \bar{t}_{0,[1]} \left( u - \ft{i}{2} \right) \right]
\widetilde{Q}^{(3)}_0 (u)
\\
- \,
u^L
t_{0,[1]} \left( u - \ft{i}{2} \right)
\!\!\!\!&&\!\!\!\!\!
\left[ \bar{t}_0^{\{2\}} (u) - \bar{t}_{0,[1]} \left( u + \ft{i}{2} \right) \bar{t}_{0,[1]} \left( u - \ft{i}{2} \right) \right]
\widetilde{Q}^{(3)}_0 (u + i)
\nonumber\\
- \,
u^L
\bar{t}_{0,[1]} \left( u + \ft{i}{2} \right)
\!\!\!\!&&\!\!\!\!\!
\left[ t_0^{\{2\}} (u) - t_{0,[1]} \left( u + \ft{i}{2} \right) t_{0,[1]} \left( u - \ft{i}{2} \right) \right]
\widetilde{Q}^{(3)}_0 (u - i)
= 0
\, . \nonumber
\ea
The similarity of Baxter equations for Baxter polynomials in terms of symmetric and antisymmetric transfer
matrices implies that there a consistency relations between them
\ba
\bar{t}_{0,[2]} (u)  t_{0,[1]} \left( u + \ft{i}{2} \right) t_{0,[1]} \left( u - \ft{i}{2} \right)
\!\!\!&=&\!\!\!
t_{0,[2]} (u)  \bar{t}_{0,[1]} \left( u + \ft{i}{2} \right) \bar{t}_{0,[1]} \left( u - \ft{i}{2} \right)
\, , \\
\bar{t}_0^{\{2\}} (u)  t_{0,[1]} \left( u + \ft{i}{2} \right) t_{0,[1]} \left( u - \ft{i}{2} \right)
\!\!\!&=&\!\!\!
t_0^{\{2\}} (u)  \bar{t}_{0,[1]} \left( u + \ft{i}{2} \right) \bar{t}_{0,[1]} \left( u - \ft{i}{2} \right)
\, .
\ea
The validity of these equations can be explicitly verified using Eqs.\ \re{Fundamental1Tranfer}.

\subsection{Particle-hole transformation}
\label{ParticleHoleTranformation}

The Bethe equations \re{DistinguishedBAElo} in the distinguished basis are not particularly convenient
for generalization beyond leading order of perturbation theory in maximally supersymmetric gauge
theory since the corresponding pseudovacuum state is not protected from quantum corrections in 't
Hooft coupling constant. Therefore, it is necessary to transform Bethe and Baxter equations to the
basis with protected pseudovacuum state (see Appendix \ref{AppendixExcitation}). For the underlying
superalgebra this reflects non-uniqueness in the choice of the simple root system. The inequivalent
root systems are related by Weyl group of super-reflections $SW (G)$ with respect to odd roots of the
superalgebra \cite{Kac77,GroLei01,FraSorSci96}, as discussed in Appendix \ref{SerreChevalleyAppendix}.
In terms of Bethe Ansatz equations this is known as the particle-hole transformation
\cite{Woy83,EssKor91,EssKorSch92,EssKor92,FoeKar92,GoeSee03,BeiSta05,BeiKazSakZar05}.

Let us perform a chain of these transformations on the distinguished Kac-Dynkin diagram yielding
non-distinguished one with FBBF grading, Fig.\ \ref{dynkin}, the central node of which will be the
$sl(2)$ subalgebra corresponding to scalar Wilson operators with covariant derivatives. At first
step, we reflect the diagram with respect to the odd root $\bit{\alpha}_2$ which translates into a
duality transformation with respect to the fermionic Bethe root $\tilde{u}^{(2)}_{0,k}$. At first,
one rewrites the Bethe equation for $\tilde{u}^{(2)}_{0,k}$ as zeros of the polynomial $P_2 (u)$ at
the positions of these roots,
\ba
0
\!\!\!&=&\!\!\!
P_2 (\tilde{u}^{(2)}_{0,k})
\\
&=&\!\!\!
\left( \tilde{u}^{(2)}_{0,k} + \ft{i}{2} \right)^L
\widetilde{Q}^{(1)}_0 \left( \tilde{u}^{(2)}_{0,k} + \ft{i}{2} \right)
\widetilde{Q}^{(3)}_0 \left( \tilde{u}^{(2)}_{0,k} - \ft{i}{2} \right)
-
\left( \tilde{u}^{(2)}_{0,k} - \ft{i}{2} \right)^L
\widetilde{Q}^{(1)}_0 \left( \tilde{u}^{(2)}_{0,k} - \ft{i}{2} \right)
\widetilde{Q}^{(3)}_0 \left( \tilde{u}^{(2)}_{0,k} + \ft{i}{2} \right)
\, . \nonumber
\ea
Obviously, one can extract the polynomial $\widetilde{Q}^{(2)}_0 (u)$ from $P_2 (u)$ with remaining
roots encoded into yet another polynomial $Q^{(2)}_0 (u)$, such that
\be
\label{ParticleHoleU2}
P_2 (u) = \Lambda_2 \widetilde{Q}^{(2)}_0 (u) Q^{(2)}_0 (u)
\, ,
\ee
where $\Lambda_2 = i (L - \tilde{n}_3 + \tilde{n}_1)$. The number $n_2$ of the dual roots $u^{(2)}_{0,k}$
of the polynomial $Q^{(2)}_0 (u)$ is related to the ones of the other Baxter functions as $n_2 = L +
\tilde{n}_1 - \tilde{n}_2 + \tilde{n}_3 - 1$. Using Eq.\ \re{ParticleHoleU2}, one eliminates the polynomial
$\widetilde{Q}^{(2)}_0 (u)$ from the Bethe equations \re{DistinguishedBAElo} and deduces Bethe equations
corresponding to the Cartan matrix
\be
A =
\left(
\begin{array}{rrr}
0  & 1  &  0 \\
1  &  0 & -1 \\
0  & -1 &  0
\end{array}
\right)
\, ,
\ee
with the (middle) Kac-Dynkin diagram in Fig.\ \ref{dynkin}. They read
\ba
\label{FermionicABAlo}
1
\!\!\!&=&\!\!\!
\frac{
Q^{(2)}_0 \left( \tilde{u}^{(1)}_{0,k} + \ft{i}{2} \right)
}{
Q^{(2)}_0 \left( \tilde{u}^{(1)}_{0,k} - \ft{i}{2} \right)
}
\, , \nonumber\\
\left(
\frac{u^{(2)}_{0,k} - \ft{i}{2}}{u^{(2)}_{0,k} + \ft{i}{2}}
\right)^L
\!\!\!&=&\!\!\!
\frac{
\widetilde{Q}^{(1)}_0 \left( u^{(2)}_{0,k} + \ft{i}{2} \right)
}{
\widetilde{Q}^{(1)}_0 \left( u^{(2)}_{0,k} - \ft{i}{2} \right)
}
\frac{
\widetilde{Q}^{(3)}_0 \left( u^{(2)}_{0,k} - \ft{i}{2} \right)
}{
\widetilde{Q}^{(3)}_0 \left( u^{(2)}_{0,k} + \ft{i}{2} \right)
}
\, , \\
1
\!\!\!&=&\!\!\!
\frac{
Q^{(2)}_0 \left( \tilde{u}^{(3)}_{0,k} - \ft{i}{2} \right)
}{
Q^{(2)}_0 \left( \tilde{u}^{(3)}_{0,k} + \ft{i}{2} \right)
}
\, . \nonumber
\ea

At the next step, we reflect the root system respect to the odd root $\bit{\alpha}_1$ of the middle
Kac-Dynkin diagram in Fig.\ \ref{dynkin}. To dualize the corresponding fermionic Bethe roots
$\tilde{u}^{(1)}_{0,k}$, we introduce yet another polynomial $P_1 (u)$ that vanishes, according to
the Bethe Ansatz equations \re{FermionicABAlo}, at $u = \tilde{u}^{(1)}_{0,k}$,
\be
0 = P_1 ( \tilde{u}^{(1)}_{0,k} )
=
Q^{(2)}_0 \left( \tilde{u}^{(1)}_{0,k} + \ft{i}{2} \right)
-
Q^{(2)}_0 \left( \tilde{u}^{(1)}_{0,k} - \ft{i}{2} \right)
\, .
\ee
The Bethe roots $\tilde{u}^{(1)}_{0,k}$ do not exhaust all zeros of the polynomial $P_1 (u)$ and,
therefore, for arbitrary $u$ it can be rewritten as a product of two polynomials
\be
P_1 (u) = \Lambda_1 \widetilde{Q}^{(1)}_0 (u) Q^{(1)}_0 (u)
\, ,
\ee
with the second one being the dual Baxter polynomial or order $n_1$ in new Bethe roots $u^{(1)}_{0,k}$.
The proportionality factor $\Lambda_1$ and the power $n_1$ are related to the numbers of ``parent" Bethe
roots as follows $\Lambda_1 = i n_2$ and $n_1 = n_2 - \tilde{n}_1 - 1$, respectively. Again, eliminating
the polynomial $\widetilde{Q}^{(1)}_0 (u)$, one gets Bethe equations for the symmetric Kac-Dynkin diagram
with two isotropic fermionic roots in (right-most) Fig.\ \ref{dynkin}, and the Cartan matrix
\be
\label{CartanSymmetric}
A =
\left(
\begin{array}{rrr}
0  &  -1 &  0 \\
-1  & 2 &  -1 \\
0  &  -1 &  0
\end{array}
\right)
\, ,
\ee
which read
\ba
\label{SymmetricABAlo}
1 \!\!\!&=&\!\!\!
\frac{
Q^{(2)}_0 \left( u^{(1)}_{0,k} - \ft{i}{2} \right)
}{
Q^{(2)}_0 \left( u^{(1)}_{0,k} + \ft{i}{2} \right)
}
\, , \nonumber\\
- \left( \frac{u^{(2)}_{0,k} - \ft{i}{2}}{u^{(2)}_{0,k} + \ft{i}{2}} \right)^L
\!\!\!&=&\!\!\!
\frac{Q^{(1)}_0 \left( u^{(2)}_{0,k} - \ft{i}{2} \right)}{Q^{(1)}_0 \left( u^{(2)}_{0,k} + \ft{i}{2} \right)}
\frac{Q^{(2)}_0 (u^{(2)}_{0,k} + i)}{Q^{(2)}_0 (u^{(2)}_{0,k} - i)}
\frac{Q^{(3)}_0 \left( u^{(2)}_{0,k} - \ft{i}{2} \right)}{Q^{(3)}_0 \left( u^{(2)}_{0,k} + \ft{i}{2} \right)}
\, , \\
1 \!\!\!&=&\!\!\!
\frac{
Q^{(2)}_0 \left( u^{(3)}_{0,k} - \ft{i}{2} \right)
}{
Q^{(2)}_0 \left( u^{(3)}_{0,k} + \ft{i}{2} \right)
}
\, , \nonumber
\ea
where for conformity of notations we renamed $\widetilde{Q}^{(3)}_0 (u) = Q^{(3)}_0 (u)$ and
$\tilde{u}^{(3)}_{0,k} = u^{(3)}_{0,k}$. In what follows, we will dub for brevity corresponding
basis symmetric.

\subsection{Transfer matrices in symmetric basis}
\label{TransferMatricesSymmetric}

Let us now construct eigenvalues of transfer matrices in symmetric basis labelled a skew Young
supertableaux \cite{Che86,Ful97}. Analogously to \re{DistinguishedBox} we identify the elementary
Young supertableaux with a product of ratios of Baxter polynomials
\ba
\label{ElementaryBoxSymmetric}
\unitlength0.4cm
\begin{picture}(2.3,1.2)
\linethickness{0.4mm}
\put(1,0){\framebox(1,1){$1$}}
\end{picture}
{}_{u}
\!\!\!&=&\!\!\!
\left(u + \ft{i}{2}\right)^L
\frac{Q^{(1)}_0 \left( u - \ft{i}{2} \right)}{Q^{(1)}_0 \left( u + \ft{i}{2} \right)}
\, , \\
\unitlength0.4cm
\begin{picture}(2.3,1)
\linethickness{0.4mm}
\put(1,0){\framebox(1,1){$2$}}
\end{picture}
{}_{u}
\!\!\!&=&\!\!\!
\left(u + \ft{i}{2}\right)^L
\frac{Q^{(1)}_0 \left( u - \ft{i}{2} \right)}{Q^{(1)}_0 \left( u + \ft{i}{2} \right)}
\frac{Q^{(2)}_0 (u + i)}{Q^{(2)}_0 (u)}
\, , \nonumber\\
\unitlength0.4cm
\begin{picture}(2.3,1)
\linethickness{0.4mm}
\put(1,0){\framebox(1,1){$3$}}
\end{picture}
{}_{u}
\!\!\!&=&\!\!\!
\left(u - \ft{i}{2}\right)^L
\frac{Q^{(3)}_0 \left( u + \ft{i}{2} \right)}{Q^{(3)}_0 \left( u - \ft{i}{2} \right)}
\frac{Q^{(2)}_0 (u - i)}{Q^{(2)}_0 (u)}
\, , \nonumber\\
\unitlength0.4cm
\begin{picture}(2.3,1)
\linethickness{0.4mm}
\put(1,0){\framebox(1,1){$4$}}
\end{picture}
{}_{u}
\!\!\!&=&\!\!\!
\left(u - \ft{i}{2}\right)^L
\frac{Q^{(3)}_0 \left( u + \ft{i}{2} \right)}{Q^{(3)}_0 \left( u - \ft{i}{2} \right)}
\, , \nonumber
\ea
with gradings $\bar{1} = \bar{4} = 1$ and $\bar{2} = \bar{3} = 0$ associated with (right-most)
Kac-Dynkin in Fig.\ \ref{dynkin}. Introducing a symbolic notation for the above single-box Young
supertableau with a flavor index $\alpha$ and labelled by the spectral parameter $u$
\be
\unitlength0.4cm
\begin{picture}(2.3,1.2)
\linethickness{0.4mm}
\put(1,0){\framebox(1,1){$\alpha$}}
\end{picture}
{}_{u}
= \,
\mathcal{Y}_0 (\alpha, u)
\, ,
\ee
we can write generating functions \cite{BazRes89,KunSuz95,KriLipWieZab97,Tsu97} for eigenvalues of
transfer matrices in antisymmetric $(1^a)$ representation
\ba
\Big[1 + \mathcal{Y}_0 (4, u) {\rm e}^{-i \partial_u}\Big]^{- 1}
\Big[1 + \mathcal{Y}_0 (3, u) {\rm e}^{-i \partial_u}\Big]
\Big[1 + \mathcal{Y}_0 (2, u) {\rm e}^{-i \partial_u}\Big]
\Big[1 \!\!\!&+&\!\!\! \mathcal{Y}_0 (1, u) {\rm e}^{-i \partial_u}\Big]^{- 1}
\\
&=&\!\!\!
\sum_{a = 0}^\infty
t_{0, [a]} \left( u - i \frac{a - 1}{2} \right)
{\rm e}^{- i a \partial_u}
\, , \nonumber
\ea
where the powers $p_\alpha = 1 - 2 \bar{\alpha}$ of factors in the left-hand side reflect the grading
of the Kac-Dynkin diagram; for symmetric $(s)$ representation one finds
\ba
\Big[1 - \mathcal{Y}_0 (1, u) {\rm e}^{-i \partial_u}\Big]
\Big[1 - \mathcal{Y}_0 (2, u) {\rm e}^{-i \partial_u}\Big]^{- 1}
\Big[1 - \mathcal{Y}_0 (3, u) {\rm e}^{-i \partial_u}\Big]^{- 1}
\Big[1 \!\!\!&-&\!\!\! \mathcal{Y}_0 (4, u) {\rm e}^{-i \partial_u}\Big]
\\
&=&\!\!\!
\sum_{s = 0}^\infty
t_0^{\{s\}} \left( u - i \frac{s - 1}{2} \right)
{\rm e}^{- i s \partial_u}
\, . \nonumber
\ea
Bethe Ansatz equations \re{SymmetricABAlo} imply that these transfer matrices are pole-free.

One can define transfer matrices with auxiliary space labelled by a skew Young superdiagram
$Y (\bit{m}/\bit{n})$ \cite{KirRes87,BazRes89}. $Y (\bit{m}/\bit{n})$ is obtained by removing a
Young superdiagram $Y (\bit{n})$ determined by the partitioning $\bit{n} = \{ n_1, n_2, \dots \}$
with the usual ordering condition on its elements $n_1 \geq n_2 \geq \dots$ from a larger superdiagram
$Y (\bit{m})$ with $\bit{m} = \{ m_1, m_2, \dots \}$ and $m_1 \geq m_2 \geq \dots$ such that $\bit{m}
\succ \bit{n}$. One enumerates all boxes of the Young superdiagram $Y (\bit{m})$ starting with the
top left one with a pair of integers $(j,k)$, $j$ and $k$ enumerating rows and columns, respectively.
On the skew superdiagram we define a set of admissible skew Young supertableaux $Y_\alpha (\bit{m}/
\bit{n})$ by assigning a flavor $\alpha (j,k)$ index to each box of the diagram $Y (\bit{m}/\bit{n})$
and distributing them according to the following ordering conditions: $\alpha (j, k) < \alpha
(j, k + 1)$ and $\alpha (j, k) < \alpha (j + 1, k)$ for any two adjacent boxes, with weaker
conditions when these indices have coincident gradings, namely, for
\begin{itemize}

\item bosonic grading $\bar\alpha = 0$:
\be
\alpha (j, k) \leq \alpha (j, k + 1)
\, , \qquad
\alpha (j, k) < \alpha (j + 1, k)
\, ;
\ee

\item fermionic grading $\bar\alpha = 1$:
\be
\alpha (j, k) < \alpha (j, k + 1)
\, , \qquad
\alpha (j, k) \leq \alpha (j + 1, k)
\, .
\ee

\end{itemize}
Obviously these flavor indices can take four different values, i.e., $1 \leq \alpha \leq 4$
for $psl(2|2)$. Notice that a Young dsuperiagram defined by the partition $\bit{m} = \{ m_1, m_2,
\dots, m_M \}$ can be equivalently represented as $Y(\bit{m}) = (s_1^{a_1}, s_2^{a_2}, \dots,
s_\ell^{a_\ell})$ with $a_1 + a_2 + \dots + a_\ell = M$ in case there are coincident $m_k$'s,
i.e., $s_1 = m_1 = \dots = m_{a_1}$, $s_2 = m_{a_1 + 1} = \dots = m_{a_2}$, ... A transposed
Young superdiagram is then obtained by reflection across the main diagonal of horizontal and
vertical rows. It can be written as $Y(\tilde{\bit{m}}) = ( (a_1 + \dots + a_\ell)^{s_\ell}, (a_1 +
\dots + a_{\ell - 1})^{s_{\ell - 1} - s_\ell}, \dots )$ where $\tilde{\bit{m}} = \{ \tilde{m}_1,
\tilde{m}_2, \dots \}$ such that $\tilde{m}_1 = M$ is the hight of the first column of
$Y(\bit{m})$, etc.

\begin{figure}[t]
\begin{center}
\mbox{
\begin{picture}(0,170)(110,0)
\put(0,0){\insertfig{8}{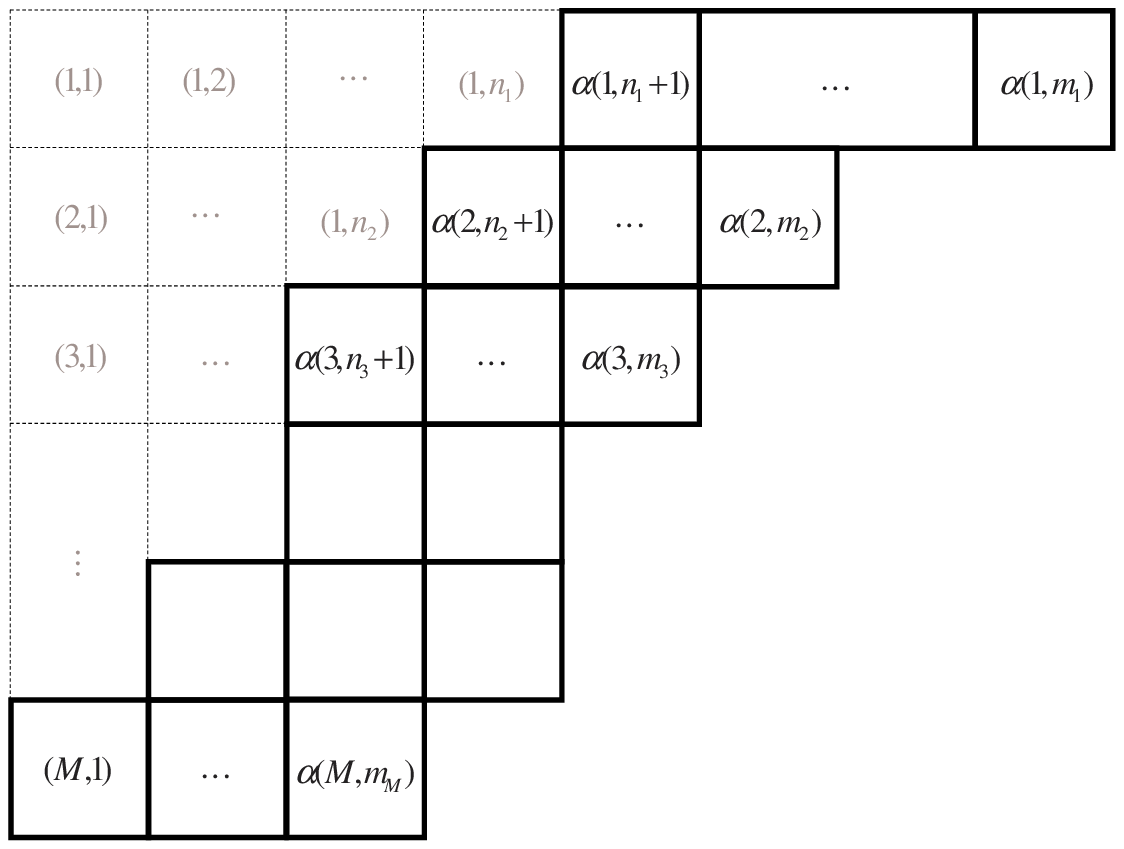}}
\end{picture}
}
\end{center}
\caption{\label{SkewYoungTableau} Skew Young supertableau $Y_\alpha (\bit{m}/\bit{n})$.}
\end{figure}

For the auxiliary space determined by a skew Young supertableau as in Fig.\ \ref{SkewYoungTableau},
the transfer matrix can be constructed from the elementary boxes \re{ElementaryBoxSymmetric},
\be
\label{SkewYoungTransferMatrix}
t_{0, Y (\bit{\scriptstyle m}/\bit{\scriptstyle n})} (u)
=
\sum_{Y_\alpha} \prod_{\alpha (j,k) \in Y_\alpha}
p_{\alpha (j,k)} \mathcal{Y}_0 \left( \alpha (j, k), u + \ft{i}{2} (\tilde{m}_1 - m_1 + 2j - 2k) \right)
\, ,
\ee
where $\tilde{m}_1 = M$. These transfer matrices are functionally dependent. They satisfy a number
of functional relations known as fusion relations, namely, they admit a determinant representation
\cite{BazRes89,KunSuz95,KunOhtSuz95,Tsu97} in terms of (anti-)symmetric transfer matrices $t_{0,[a]}$ and
$t_0^{\{s\}}$,
\ba
\label{DeterminantFusionLO}
t_{0, Y (\bit{\scriptstyle m}/\bit{\scriptstyle n})} (u)
\!\!\!&=&\!\!\!
\det_{1 \leq j, k \leq m_1}
t_{0, [\tilde{m}_j - \tilde{n}_k - j + k]}
\left(
u + \ft{i}{2}
\left( m_1 - \tilde{m}_1 + \tilde{m}_j + \tilde{n}_k - j - k + 1 \right)
\right)
\nonumber\\
&=&\!\!\!
\det_{1 \leq j, k \leq \tilde{m}_1}
t_0^{\{m_k - n_j + j - k\}}
\left(
u + \ft{i}{2}
\left( m_1 - \tilde{m}_1 - m_k - n_j + j + k - 1 \right)
\right)
\, ,
\ea
with $t_{0, [a < 0]} (u) = t_0^{\{s < 0\}} (u) = 0$. For instance for a skew Young superdiagram
$Y (\bit{m}'/\bit{n}')$ with $\bit{m}' = \{ 2,2,2 \}$ and $\bit{n}' = \{1, 1\}$, it yields
\ba
t_{0, Y (\bit{\scriptstyle m}'/\bit{\scriptstyle n}')} (u)
\!\!\!&=&\!\!\!
\det
\left(
\begin{array}{lll}
t_0^{\{1\}} \left( u - \ft{3 i}{2} \right)
&
1
&
0
\\
t_0^{\{2\}} (u - i)
&
t_0^{\{1\}} \left( u - \ft{i}{2} \right)
&
1
\\
t_0^{\{4\}} (u)
&
t_0^{\{3\}} \left( u + \ft{i}{2} \right)
&
t_0^{\{2\}} (u - i)
\end{array}
\right)
\\
&=&\!\!\!
\det
\left(
\begin{array}{ll}
t_{0,[1]} \left( u + \ft{3 i}{2} \right)
&
t_{0,[4]} (u)
\\
1
&
t_{0, [3]} \left( u - \ft{i}{2} \right)
\end{array}
\right)
\, . \nonumber
\ea
Equations \re{DeterminantFusionLO} are a generalization of classical formulas for characters
on representation determined by corresponding Young supertableaux \cite{BalBar81}.

Using the determinant representation for transfer matrices $t_{0, Y (\bit{\scriptstyle m}/\bit{\scriptstyle n})}
(u)$, one can immediately find bilinear fusion relations among them \cite{KluPea92,KunNakSuz94,KriLipWieZab97}.
Making use of the Desnanot-Jacobi determinant identity, one immediately finds relations for transfer matrices
with auxiliary space corresponding to rectangular Young superdiagrams $Y(\bit{m}/\varnothing) = (s^a)$, $t_{0,(s^a)}
= t_{0,[a]}^{\{s\}}$:
\be
t_{0,[a]}^{\{s\}} \left( u + \ft{i}{2} \right)
t_{0,[a]}^{\{s\}} \left( u - \ft{i}{2} \right)
=
t_{0,[a]}^{\{s + 1\}} (u) t_{0,[a]}^{\{s - 1\}} (u)
+
t_{0,[a + 1]}^{\{s\}} (u) t_{0,[a - 1]}^{\{s\}} (u)
\, .
\ee
This is a Hirota bilinear difference equations derived in Ref.\ \cite{Tsu97} and recently discussed in
\cite{KazSorZab07}.

Similarly to the distinguished basis, it turns out however that out of the entire tower of transfer
matrices, we will need just the lowest-dimensional ones
\ba
\label{t1symmetric}
t_{0,[1]} (u)
\!\!\!&=&\!\!\!
t_0^{\{1\}} (u)
=
-
\unitlength0.4cm
\begin{picture}(2.3,1)
\linethickness{0.4mm}
\put(1,-0.2){\framebox(1,1){$1$}}
\end{picture}
{}_{u}
+\!\!
\unitlength0.4cm
\begin{picture}(2.3,1)
\linethickness{0.4mm}
\put(1,-0.2){\framebox(1,1){$2$}}
\end{picture}
{}_{u}
+\!\!
\unitlength0.4cm
\begin{picture}(2.3,1)
\linethickness{0.4mm}
\put(1,-0.2){\framebox(1,1){$3$}}
\end{picture}
{}_{u}
-\!\!
\unitlength0.4cm
\begin{picture}(2.3,1)
\linethickness{0.4mm}
\put(1,-0.2){\framebox(1,1){$4$}}
\end{picture}
{}_{u}
\\[2mm]
&=&\!\!\!\!
\left(u - \ft{i}{2}\right)^L
\frac{Q^{(3)}_0 \left( u + \ft{i}{2} \right)}{Q^{(3)}_0 \left( u - \ft{i}{2} \right)}
\left(
\frac{Q^{(2)}_0 (u - i)}{Q^{(2)}_0 (u)} - 1
\right)
+
\left(u + \ft{i}{2}\right)^L
\frac{Q^{(1)}_0 \left( u - \ft{i}{2} \right)}{Q^{(1)}_0 \left( u + \ft{i}{2} \right)}
\left(
\frac{Q^{(2)}_0 (u + i)}{Q^{(2)}_0 (u)} - 1
\right)
\! , \nonumber
\ea
and
\ba
\label{t2asymmetric}
t_{0,[2]} (u)
\!\!\!&=&\!\!\!
- \!\!
\unitlength0.4cm
\begin{picture}(2.3,1)
\linethickness{0.4mm}
\put(1,0.5){\framebox(1,1){$1$}}
\put(1,-0.6){\framebox(1,1){$2$}}
\end{picture}
\!\!
{\begin{array}{c} \scriptstyle u - \frac{i}{2} \vspace{-0.5mm} \\ \scriptstyle u + \frac{i}{2} \end{array}}
\!\!-\!\!\!\!
\unitlength0.4cm
\begin{picture}(2.3,1)
\linethickness{0.4mm}
\put(1,0.5){\framebox(1,1){$1$}}
\put(1,-0.6){\framebox(1,1){$3$}}
\end{picture}
\!\!
{\begin{array}{c} \scriptstyle u - \frac{i}{2} \vspace{-0.5mm} \\ \scriptstyle u + \frac{i}{2} \end{array}}
\!\!+\!\!\!\!
\unitlength0.4cm
\begin{picture}(2.3,1)
\linethickness{0.4mm}
\put(1,0.5){\framebox(1,1){$1$}}
\put(1,-0.6){\framebox(1,1){$4$}}
\end{picture}
\!\!
{\begin{array}{c} \scriptstyle u - \frac{i}{2} \vspace{-0.5mm} \\ \scriptstyle u + \frac{i}{2} \end{array}}
\!\!+\!\!\!\!
\unitlength0.4cm
\begin{picture}(2.3,1)
\linethickness{0.4mm}
\put(1,0.5){\framebox(1,1){$2$}}
\put(1,-0.6){\framebox(1,1){$3$}}
\end{picture}
\!\!
{\begin{array}{c} \scriptstyle u - \frac{i}{2} \vspace{-0.5mm} \\ \scriptstyle u + \frac{i}{2} \end{array}}
\!\!-\!\!\!\!
\unitlength0.4cm
\begin{picture}(2.3,1)
\linethickness{0.4mm}
\put(1,0.5){\framebox(1,1){$2$}}
\put(1,-0.6){\framebox(1,1){$4$}}
\end{picture}
\!\!
{\begin{array}{c} \scriptstyle u - \frac{i}{2} \vspace{-0.5mm} \\ \scriptstyle u + \frac{i}{2} \end{array}}
\!\!-\!\!\!\!
\unitlength0.4cm
\begin{picture}(2.3,1)
\linethickness{0.4mm}
\put(1,0.5){\framebox(1,1){$3$}}
\put(1,-0.6){\framebox(1,1){$4$}}
\end{picture}
\!\!
{\begin{array}{c} \scriptstyle u - \frac{i}{2} \vspace{-0.5mm} \\ \scriptstyle u + \frac{i}{2} \end{array}}
\!\!+\!\!\!\!
\unitlength0.4cm
\begin{picture}(2.3,1)
\linethickness{0.4mm}
\put(1,0.5){\framebox(1,1){$1$}}
\put(1,-0.6){\framebox(1,1){$1$}}
\end{picture}
\!\!
{\begin{array}{c} \scriptstyle u - \frac{i}{2} \vspace{-0.5mm} \\ \scriptstyle u + \frac{i}{2} \end{array}}
\!\!+\!\!\!\!
\unitlength0.4cm
\begin{picture}(2.3,1)
\linethickness{0.4mm}
\put(1,0.5){\framebox(1,1){$4$}}
\put(1,-0.6){\framebox(1,1){$4$}}
\end{picture}
\!\!
{\begin{array}{c} \scriptstyle u - \frac{i}{2} \vspace{-0.5mm} \\ \scriptstyle u + \frac{i}{2} \end{array}}
\nonumber\\[2mm]
&=&\!\!\!
u^L (u - i)^L
\frac{Q^{(3)}_0 (u + i)}{Q^{(3)}_0 (u - i)}
\left(
1 - \frac{Q^{(2)}_0 \left( u - \ft{3 i}{2} \right)}{Q^{(2)}_0 \left( u - \ft{i}{2} \right)}
\right)
\nonumber\\
&-&\!\!\!
u^{2 L}
\frac{Q^{(1)}_0 (u - i)}{Q^{(1)}_0 (u)} \frac{Q^{(3)}_0 (u + i)}{Q^{(3)}_0 (u)}
\frac{
\left(
Q^{(2)}_0 \left( u + \ft{i}{2} \right) - Q^{(2)}_0 \left( u - \ft{i}{2} \right)
\right)^2
}{Q^{(2)}_0 \left( u + \ft{i}{2} \right) Q^{(2)}_0 \left( u - \ft{i}{2} \right)}
\\
\!\!\!&+&\!\!\!
u^L (u + i)^L
\frac{Q^{(1)}_0 (u - i)}{Q^{(1)}_0 (u + i)}
\left(
1 - \frac{Q^{(2)}_0 \left( u + \ft{3 i}{2} \right)}{Q^{(2)}_0 \left( u + \ft{i}{2} \right)}
\right)
\, , \nonumber
\ea
or
\ba
\label{t2ssymmetric}
t_0^{\{2\}} (u)
\!\!\!&=&\!\!\!
- \!\!
\unitlength0.4cm
\begin{picture}(4,1)
\linethickness{0.4mm}
\put(1,0){\framebox(1,1){$1$}}
\put(2.1,0){\framebox(1,1){$2$}}
\put(0.4,-1.3){$\scriptstyle u + \frac{i}{2}$}
\put(2.2,-1.3){$\scriptstyle u - \frac{i}{2}$}
\end{picture}
\!\!- \!\!
\unitlength0.4cm
\begin{picture}(4,1)
\linethickness{0.4mm}
\put(1,0){\framebox(1,1){$1$}}
\put(2.1,0){\framebox(1,1){$3$}}
\put(0.4,-1.3){$\scriptstyle u + \frac{i}{2}$}
\put(2.2,-1.3){$\scriptstyle u - \frac{i}{2}$}
\end{picture}
\!\!+ \!\!
\unitlength0.4cm
\begin{picture}(4,1)
\linethickness{0.4mm}
\put(1,0){\framebox(1,1){$1$}}
\put(2.1,0){\framebox(1,1){$4$}}
\put(0.4,-1.3){$\scriptstyle u + \frac{i}{2}$}
\put(2.2,-1.3){$\scriptstyle u - \frac{i}{2}$}
\end{picture}
\!\!+ \!\!
\begin{picture}(4,1)
\linethickness{0.4mm}
\put(1,0){\framebox(1,1){$2$}}
\put(2.1,0){\framebox(1,1){$3$}}
\put(0.4,-1.3){$\scriptstyle u + \frac{i}{2}$}
\put(2.2,-1.3){$\scriptstyle u - \frac{i}{2}$}
\end{picture}
\!\!- \!\!
\unitlength0.4cm
\begin{picture}(4,1)
\linethickness{0.4mm}
\put(1,0){\framebox(1,1){$2$}}
\put(2.1,0){\framebox(1,1){$4$}}
\put(0.4,-1.3){$\scriptstyle u + \frac{i}{2}$}
\put(2.2,-1.3){$\scriptstyle u - \frac{i}{2}$}
\end{picture}
\!\!- \!\!
\begin{picture}(4,1)
\linethickness{0.4mm}
\put(1,0){\framebox(1,1){$3$}}
\put(2.1,0){\framebox(1,1){$4$}}
\put(0.4,-1.3){$\scriptstyle u + \frac{i}{2}$}
\put(2.2,-1.3){$\scriptstyle u - \frac{i}{2}$}
\end{picture}
\!\!+ \!\!
\unitlength0.4cm
\begin{picture}(4,1)
\linethickness{0.4mm}
\put(1,0){\framebox(1,1){$2$}}
\put(2.1,0){\framebox(1,1){$2$}}
\put(0.4,-1.3){$\scriptstyle u + \frac{i}{2}$}
\put(2.2,-1.3){$\scriptstyle u - \frac{i}{2}$}
\end{picture}
\!\!+ \!\!
\unitlength0.4cm
\begin{picture}(4,1)
\linethickness{0.4mm}
\put(1,0){\framebox(1,1){$3$}}
\put(2.1,0){\framebox(1,1){$3$}}
\put(0.4,-1.3){$\scriptstyle u + \frac{i}{2}$}
\put(2.2,-1.3){$\scriptstyle u - \frac{i}{2}$}
\end{picture}
\nonumber\\[7mm]
&=&\!\!\!
u^L (u - i)^L
\frac{Q^{(3)}_0 (u + i)}{Q^{(3)}_0 (u - i)}
\frac{Q^{(2)}_0 \left( u - \ft{i}{2} \right)}{Q^{(2)}_0 \left( u + \ft{i}{2} \right)}
\left(
\frac{Q^{(2)}_0 \left( u - \ft{3 i}{2} \right)}{Q^{(2)}_0 \left( u - \ft{i}{2} \right)} - 1
\right)
\nonumber\\
&+&\!\!\!
(u - i)^L (u + i)^L
\frac{Q^{(3)}_0 (u)}{Q^{(3)}_0 (u - i)} \frac{Q^{(1)}_0 (u)}{Q^{(1)}_0 (u + i)}
\left(
\frac{Q^{(2)}_0 \left( u - \ft{3 i}{2} \right)}{Q^{(2)}_0 \left( u - \ft{i}{2} \right)} - 1
\right)
\left(
\frac{Q^{(2)}_0 \left( u + \ft{3 i}{2} \right)}{Q^{(2)}_0 \left( u + \ft{i}{2} \right)} - 1
\right)
\nonumber\\
&+&\!\!\!
u^L (u + i)^L
\frac{Q^{(1)}_0 (u - i)}{Q^{(1)}_0 (u + i)}
\frac{Q^{(2)}_0 \left( u + \ft{i}{2} \right)}{Q^{(2)}_0 \left( u - \ft{i}{2} \right)}
\left(
\frac{Q^{(2)}_0 \left( u + \ft{3 i}{2} \right)}{Q^{(2)}_0 \left( u + \ft{i}{2} \right)} - 1
\right)
\, ,
\ea
and their conjugate, for the derivation of Baxter equations for the polynomials $Q^{(k)}_0$.

\subsection{Baxter equations in symmetric basis}
\label{LOBaxterSymmetric}

The transfer matrices \re{t1symmetric}, \re{t2asymmetric} and \re{t2ssymmetric} and their conjugate can
be used to find closed equations for nested Baxter polynomials analogously to the distinguished basis
as we discussed in Sect.\ \ref{TransferBaxterDistinguished}.

First, solving conjugate transfer matrices for the polynomial $Q^{(2)}_0$, one immediately finds the
following relations involving both $Q^{(1)}_0$ and $Q^{(3)}_0$,
\ba
\label{BeautyQ1Q3}
Q^{(1)}_0 \left( u + \ft{i}{2} \right) Q^{(3)}_0 \left( u - \ft{i}{2} \right) t_{0,[1]} (u)
\!\!\!&=&\!\!\!
Q^{(1)}_0 \left( u - \ft{i}{2} \right) Q^{(3)}_0 \left( u + \ft{i}{2} \right) \bar{t}_{0,[1]} (u)
\, , \\
Q^{(1)}_0 (u + i) Q^{(3)}_0 (u - i) t_{0,[2]} (u)
\!\!\!&=&\!\!\!
Q^{(1)}_0 (u - i) Q^{(3)}_0 (u + i) \bar{t}_{0,[2]} (u)
\, , \\
Q^{(1)}_0 (u + i) Q^{(3)}_0 (u - i) t_0^{\{2\}} (u)
\!\!\!&=&\!\!\!
Q^{(1)}_0 (u - i) Q^{(3)}_0 (u + i) \bar{t}_0^{\{2\}} (u)
\, .
\ea
They can be further generalized for arbitrary length of Young supertableaux as shown in Eq.\
\re{Q1Q3RelationsAllOrder} for their all-order analogues. The similarity of the last two equations
is a consequence of functional relations between these transfer matrices
\ba
\label{TreeRelationsTransfer}
t_{0,[1]} \left( u + \ft{i}{2} \right)
t_{0,[1]} \left( u - \ft{i}{2} \right)
\bar{t}_{0,[2]} (u)
\!\!\!&=&\!\!\!
\bar{t}_{0,[1]} \left( u + \ft{i}{2} \right)
\bar{t}_{0,[1]} \left( u - \ft{i}{2} \right)
t_{0,[2]} (u)
\, , \\
t_{0,[1]} \left( u + \ft{i}{2} \right)
t_{0,[1]} \left( u - \ft{i}{2} \right)
\bar{t}_0^{\{2\}} (u)
\!\!\!&=&\!\!\!
\bar{t}_{0,[1]} \left( u + \ft{i}{2} \right)
\bar{t}_{0,[1]} \left( u - \ft{i}{2} \right)
t_0^{\{2\}} (u)
\, .
\ea

Now, finding $Q^{(2)}_0$ from $\bar{t}_{0,[1]}$ and eliminating it from $t_{0,[2]}$, we get
\be
\label{IntermediateBaxterQ1Q3}
t_{0,[2]} (u)
+
u^L
\bar{t}_{0,[1]} \left( u + \ft{i}{2} \right) \frac{Q^{(1)}_0 (u - i)}{Q^{(1)}_0 (u + i)} \frac{Q^{(3)}_0 (u + i)}{Q^{(3)}_0 (u)}
+
u^L
\bar{t}_{0,[1]} \left( u - \ft{i}{2} \right) \frac{Q^{(1)}_0 (u - i)}{Q^{(1)}_0 (u)} \frac{Q^{(3)}_0 (u + i)}{Q^{(3)}_0 (u - i)}
= 0
\, .
\ee
Here is when the relations \re{BeautyQ1Q3} becomes important for derivation of an autonomous
finite-difference equations. Using Eqs.\ \re{BeautyQ1Q3}, one can eliminate either $Q^{(1)}_0$ or
$Q^{(3)}_0$ from \re{IntermediateBaxterQ1Q3} and its conjugate, and obtain a form of Baxter equations
for these polynomials,
\ba
t_{0,[2]} (u) \bar{t}_{0,[2]} (u) Q_0^{(1)} (u)
\!\!\!&+&\!\!\!
u^L t_{0,[2]} (u) \bar{t}_{0,[1]} \left( u - \ft{i}{2} \right) Q_0^{(1)} (u + i)
\\
&+&\!\!\!
u^L \bar{t}_{0,[2]} (u) t_{0,[1]} \left( u + \ft{i}{2} \right) Q_0^{(1)} (u - i)
= 0
\, , \nonumber\\
t_{0,[2]} (u) \bar{t}_{0,[2]} (u) Q_0^{(3)} (u)
\!\!\!&+&\!\!\!
u^L \bar{t}_{0,[2]} (u) t_{0,[1]} \left( u - \ft{i}{2} \right) Q_0^{(3)} (u + i)
\\
&+&\!\!\!
u^L t_{0,[2]} (u) \bar{t}_{0,[1]} \left( u + \ft{i}{2} \right) Q_0^{(3)} (u - i)
= 0
\, . \nonumber
\ea
Since the Baxter function $Q_0^{(3)}$ was not affected by the series of particle-hole transformation,
its Baxter equation in symmetric and distinguished \re{BaxterQ3tildeDistinguished} bases coincide.

Similar Baxter equations can be derived using the symmetric transfer matrix $t_0^{\{2\}}$. Solving the
transfer matrices $t_{0,[1]}$ and $t_0^{\{2\}}$ with respect to $Q_0^{(2)}$, one finds
\be
t_{0,[1]} \left( u - \ft{i}{2} \right)
t_{0,[1]} \left( u + \ft{i}{2} \right)
-
t_0^{\{2\}} (u)
+
u^L
t_{0,[1]} \left( u + \ft{i}{2} \right)
\frac{Q^{(1)}_0 (u - i)}{Q^{(1)}_0 (u)}
+
u^L
t_{0,[1]} \left( u - \ft{i}{2} \right)
\frac{Q^{(3)}_0 (u + i)}{Q^{(3)}_0 (u)}
=
0
\, .
\ee
Deriving a similar equation for conjugate transfer matrices and solving the resulting system with respect
to either $Q_0^{(1)}$ or $Q_0^{(3)}$, one deduces two Baxter equations
\ba
\left[ t_0^{\{2\}} (u) - t_{0,[1]} \left( u + \ft{i}{2} \right) t_{0,[1]} \left( u - \ft{i}{2} \right) \right]
\!\!\!\!&&\!\!\!\!\!
\left[ \bar{t}_0^{\{2\}} (u) - \bar{t}_{0,[1]} \left( u + \ft{i}{2} \right) \bar{t}_{0,[1]} \left( u - \ft{i}{2} \right) \right]
Q^{(1)}_0 (u)
\\
- \,
u^L
\bar{t}_{0,[1]} \left( u - \ft{i}{2} \right)
\!\!\!\!&&\!\!\!\!\!
\left[ t_0^{\{2\}} (u) - t_{0,[1]} \left( u + \ft{i}{2} \right) t_{0,[1]} \left( u - \ft{i}{2} \right) \right]
Q^{(1)}_0 (u + i)
\nonumber\\
- \,
u^L
t_{0,[1]} \left( u + \ft{i}{2} \right)
\!\!\!\!&&\!\!\!\!\!
\left[ \bar{t}_0^{\{2\}} (u) - \bar{t}_{0,[1]} \left( u + \ft{i}{2} \right) \bar{t}_{0,[1]} \left( u - \ft{i}{2} \right) \right]
Q^{(1)}_0 (u - i)
= 0
\, , \nonumber\\
\left[ t_0^{\{2\}} (u) - t_{0,[1]} \left( u + \ft{i}{2} \right) t_{0,[1]} \left( u - \ft{i}{2} \right) \right]
\!\!\!\!&&\!\!\!\!\!
\left[ \bar{t}_0^{\{2\}} (u) - \bar{t}_{0,[1]} \left( u + \ft{i}{2} \right) \bar{t}_{0,[1]} \left( u - \ft{i}{2} \right) \right]
Q^{(3)}_0 (u)
\\
- \,
u^L
t_{0,[1]} \left( u - \ft{i}{2} \right)
\!\!\!\!&&\!\!\!\!\!
\left[ \bar{t}_0^{\{2\}} (u) - \bar{t}_{0,[1]} \left( u + \ft{i}{2} \right) \bar{t}_{0,[1]} \left( u - \ft{i}{2} \right) \right]
Q^{(3)}_0 (u + i)
\nonumber\\
- \,
u^L
\bar{t}_{0,[1]} \left( u + \ft{i}{2} \right)
\!\!\!\!&&\!\!\!\!\!
\left[ t_0^{\{2\}} (u) - t_{0,[1]} \left( u + \ft{i}{2} \right) t_{0,[1]} \left( u - \ft{i}{2} \right) \right]
Q^{(3)}_0 (u - i)
= 0
\, . \nonumber
\ea
Again, the equations for $Q_0^{(3)}$ in the distinguished and symmetric bases coincide by the same
token as before. Transfer matrices are polynomials in spectral parameter with coefficients determined
by local conserved charges of the spin chain. The above equations provide quantization conditions on
these charges and determine the Baxter polynomials $Q_0^{(1,3)}$. This data is used in turn to find the
momentum-carrying roots of $Q_0^{(2)}$ via Eq.\ \re{t1symmetric}. The latter determines one-loop
anomalous dimensions of superconformal harmonics in the decomposition of the composite light-cone
operators \re{LCoperator}.

\section{Long-range spin chain}
\label{LongRangeSpinChain}

Now we turn to multiloop generalization of Bethe Ansatz equations which describe the spectrum of anomalous
dimensions of Wilson operators in $sl(2|2)$ sector of $\mathcal{N} = 4$ super-Yang-Mills theory to all-orders
in 't Hooft coupling $g = g_{\rm\scriptscriptstyle YM} \sqrt{N_c}/(2 \pi)$. As we pointed out in Sect.\
\ref{ParticleHoleTranformation}, the deformation one-loop equations beyond leading order of perturbation
theory is achieved in the symmetric basis \re{SymmetricABAlo}. There are several important changes which
occur when the long-range effects enter the game. The long-rage spin chain is written in terms of
the renormalized spectral parameter \cite{BeiDipSta04}
\be
x[u] = \ft{1}{2}\left( u + \sqrt{u^2 - g^2} \right)
\, .
\ee
The scattering matrix of momentum-carrying excitations in long-range Bethe equation in the symmetric
basis acquire an additional phase factor due to renormalization of the superconformal spin and also a
nontrivial scattering phase absorbing deviations from strong-coupling calculations. Then the conjectured
form of all-order $psl(2|2)$ asymptotic Bethe equations read \cite{BeiSta05}
\ba
\label{AllONABA}
1
\!\!\!&=&\!\!\!
\prod_{j = 1}^{n_2}
\frac{x_k^{(1)} - x_j^{(2)+}}{x_k^{(1)} - x_j^{(2)-}}
\, , \\
\left( \frac{x_k^{(2)+}}{x_k^{(2)-}} \right)^L
\!\!\!&=&\!\!\!
\prod_{j \neq k = 1}^{n_2} \frac{x_k^{(2)-} - x_j^{(2)+}}{x_k^{(2)+} - x_j^{(2)-}}
\frac{
\Big( 1 - \frac{g^2}{4 x_k^{(2)+} x_j^{(2)-}} \Big)
}{
\Big( 1 - \frac{g^2}{4 x_k^{(2)-} x_j^{(2)+}} \Big)
}
\frac{
{\rm e}^{i \theta \left( x^{(2)+}_k, x^{(2)}_j \right)}
}{
{\rm e}^{i \theta \left( x^{(2)-}_k, x^{(2)}_j \right)}
}
\prod_{m = 1}^{n_1}
\frac{x_k^{(2)+} - x_m^{(1)}}{x_k^{(2)-} - x_m^{(1)}}
\prod_{n = 1}^{n_3}
\frac{x_k^{(2)+} - x_n^{(3)}}{x_k^{(2)-} - x_n^{(3)}}
\, , \nonumber\\
1
\!\!\!&=&\!\!\!
\prod_{j = 1}^{n_2}
\frac{x_k^{(3)} - x_j^{(2)+}}{x_k^{(3)} - x_j^{(2)-}}
\, . \nonumber
\ea
The precise form of the dressing factor $\theta (x_j, x_k)$ \cite{BeiEdeSta06} will be irrelevant for
the present study since the construction is purely algebraic and relies on perturbative analyticity. As
in previous discussion we will assume the approach based on conjectured form of nested Bethe Ansatz
equations and subsequent use of the analytic Bethe Ansatz to find transfer matrices. The cancellation
of the pole will be done only in perturbative, asymptotic sense, thus neglecting poles generated at
finite coupling.

\subsection{Transfer matrices}

To start with, we define three Baxter polynomials with zeros determined by three sets of Bethe roots
\ba
\label{BaxterAllOrder}
Q^{(p)} (u) = \prod_{k = 1}^{n_p} \left( u - u_k^{(p)} (g) \right)
\, .
\ea
The roots $u_k^{(p)} (g)$ depend on the 't Hooft coupling and admit an infinite perturbative expansion
\be
u_k^{(p)} (g) = u_{0,k}^{(p)} + g^2 u_{1,k}^{(p)} + \dots
\, ,
\ee
with the lowest order term being the one-loop Bethe roots $u_{0,k}^{(p)}$ obeying the Bethe Ansatz
equations \re{SymmetricABAlo}. Following our one-loop considerations in preceding sections, the eigenvalues
of transfer matrices will be built in terms of elements which parameterize components of a Young supertableaux
associated with an auxiliary space. The single-box Young supertableaux, depending on the spectral parameter $u$
and labelled by the flavor index $\alpha = 1, {\dots}, 4$ with gradings $\bar{1} = \bar{4} = 1$ and
$\bar{2} = \bar{3} = 0$, read
\ba
\label{SymmetricBoxAllOrder}
\unitlength0.4cm
\begin{picture}(2.3,1.2)
\linethickness{0.4mm}
\put(1,0){\framebox(1,1){$1$}}
\end{picture}
{}_{u}
\!\!\!&=&\!\!\!
(x^+)^L
{\rm e}^{\frac{1}{2} \Delta_+ (x^+) + \frac{1}{2} \Delta_- (x^+)}
\frac{\widehat{Q}^{(1)} \left( u - \ft{i}{2}\right)}{\widehat{Q}^{(1)} \left( u + \ft{i}{2}\right)}
\, , \\
\unitlength0.4cm
\begin{picture}(2.3,1)
\linethickness{0.4mm}
\put(1,0){\framebox(1,1){$2$}}
\end{picture}
{}_{u}
\!\!\!&=&\!\!\!
(x^+)^L
{\rm e}^{\Delta_+ (x^+)}
\frac{Q^{(2)} (u + i)}{Q^{(2)} (u)}
\frac{\widehat{Q}^{(1)} \left( u - \ft{i}{2}\right)}{\widehat{Q}^{(1)} \left( u + \ft{i}{2}\right)}
\, , \nonumber\\
\unitlength0.4cm
\begin{picture}(2.3,1)
\linethickness{0.4mm}
\put(1,0){\framebox(1,1){$3$}}
\end{picture}
{}_{u}
\!\!\!&=&\!\!\!
(x^-)^L
{\rm e}^{\Delta_- (x^-)}
\frac{Q^{(2)} (u - i)}{Q^{(2)} (u)}
\frac{\widehat{Q}^{(3)} \left( u + \ft{i}{2}\right)}{\widehat{Q}^{(3)} \left( u - \ft{i}{2}\right)}
\, , \nonumber\\
\unitlength0.4cm
\begin{picture}(2.3,1)
\linethickness{0.4mm}
\put(1,0){\framebox(1,1){$4$}}
\end{picture}
{}_{u}
\!\!\!&=&\!\!\!
(x^-)^L
{\rm e}^{\frac{1}{2} \Delta_+ (x^-) + \frac{1}{2} \Delta_- (x^-)}
\frac{\widehat{Q}^{(3)} \left( u + \ft{i}{2}\right)}{\widehat{Q}^{(3)} \left( u - \ft{i}{2}\right)}
\, . \nonumber
\ea
They depend on the dressing factor
\be
\Delta_\pm (x) = \sigma^{(2)}_\pm (x) - \Theta (x)
\, ,
\ee
which is composed of a ``trivial'' one \cite{Bel06}
\be
\label{AllLoopSigma}
\sigma^{(p)}_\eta (x)
=
\int_{- 1}^1 \frac{dt}{\pi} \,
\frac{\ln Q^{(p)} \left( \eta \ft{i}{2} - g t \right)}{\sqrt{1 - t^2}}
\left(
1 - \frac{\sqrt{u^2 - g^2}}{u + g t}
\right)
\, ,
\ee
roughly accounting for renormalization of the superconformal spin in higher orders of perturbation
theory, as explained in Ref.\ \cite{BelKorMul06}, and a ``nontrivial'' scattering factor corresponding
to $\theta (x_j, x_k)$ in long-range Bethe Ansatz equations admitting an integral form in terms of the
nested Baxter
polynomials \cite{Bel07}
\ba
\Theta (x)
\!\!\!&=&\!\!\!
g \int_{- 1}^1 \frac{d t}{\sqrt{1 - t^2}}
\ln \frac{Q^{(2)} \left( - \ft{i}{2} - g t \right)}{Q^{(2)} \left( \ft{i}{2} - g t \right)}
\,
{-\!\!\!\!\!\!\int}_{-1}^1 ds \frac{\sqrt{1 - s^2}}{s - t}
\nonumber\\
&\times&
\int_{C_{[i, i \infty]}} \frac{d \kappa}{2 \pi i} \frac{1}{\sinh^2 (\pi \kappa)} \
\ln
\left( 1 + \frac{g^2}{4 x x [\kappa + g s]} \right)
\left( 1 - \frac{g^2}{4 x x [\kappa - g s]} \right)
\, ,
\ea
cf.\ Ref.\ \cite{DorHofMal07}.  Here we also used a notation for a product of the nested Baxter functions
and the ``trivial'' dressing which emerges in all formulas
\be
\widehat{Q}^{(p)} (u) = {\rm e}^{\frac{1}{2} \sigma_0^{(p)} (x [u])} Q^{(p)} (u)
\, ,
\ee
for $p = 1,3$. For vanishing gauge coupling $g = 0$, Eqs.\ \re{SymmetricBoxAllOrder} reduce to
\re{ElementaryBoxSymmetric}.

Introducing a symbolic notation for the elementary Young supertableaux \re{SymmetricBoxAllOrder}
\be
\unitlength0.4cm
\begin{picture}(2.3,1.2)
\linethickness{0.4mm}
\put(1,0){\framebox(1,1){$\alpha$}}
\end{picture}
{}_{u}
= \,
\mathcal{Y} (\alpha, u)
\, ,
\ee
one can write generating functions for transfer matrices with antisymmetric $(1^a)$
\ba
\label{GFantiAllOrder}
\Big[1 + \mathcal{Y} (4, u) {\rm e}^{-i \partial_u}\Big]^{- 1}
\Big[1 + \mathcal{Y} (3, u) {\rm e}^{-i \partial_u}\Big]
\Big[1 + \mathcal{Y} (2, u) {\rm e}^{-i \partial_u}\Big]
\Big[1 \!\!\!&+&\!\!\! \mathcal{Y} (1, u) {\rm e}^{-i \partial_u}\Big]^{- 1}
\\
&=&\!\!\!
\sum_{a = 0}^\infty
t_{[a]} \left( x^{[1 - a]} \right)
{\rm e}^{- i a \partial_u}
\, , \nonumber
\ea
and symmetric $(s)$ finite-dimensional chiral representations,
\ba
\label{GFsymmAllOrder}
\Big[1 - \mathcal{Y} (1, u) {\rm e}^{-i \partial_u}\Big]
\Big[1 - \mathcal{Y} (2, u) {\rm e}^{-i \partial_u}\Big]^{- 1}
\Big[1 - \mathcal{Y} (3, u) {\rm e}^{-i \partial_u}\Big]^{- 1}
\Big[1 \!\!\!&-&\!\!\! \mathcal{Y} (4, u) {\rm e}^{-i \partial_u}\Big]
\\
&=&\!\!\!
\sum_{s = 0}^\infty
t^{\{s\}} \left( x^{[1 - s]} \right)
{\rm e}^{- i s \partial_u}
\, . \nonumber
\ea
In order to write transfer matrices in a concise fashion, we used
\be
u^{[\pm n]} \equiv u \pm \ft{i}{2} n
\, , \qquad
x^{[\pm n]} \equiv x \left( u^{[\pm n]} \right)
\, .
\ee
The transfer matrices $t_{[s]}$ and $t^{\{s\}}$ are free from poles upon the use of Bethe
Ansatz equations. Notice that in our treatment we assume perturbative analyticity in the
spectral parameter $u$, ignoring dynamical pole generated at finite coupling constant.
The incorporation of these into the analysis --- a problem which was recently addressed
in a related context in Ref.\ \cite{Bei06} --- goes beyond the scope of this paper.

The eigenvalues of transfer matrices with auxiliary space determined by a skew Young supertableau
in Fig.\ \ref{SkewYoungTableau} are build from the elementary boxes \re{SymmetricBoxAllOrder}
using the same algorithm as spelled out in Sect.\ \ref{TransferMatricesSymmetric} and takes the
same functional form as Eq.\ \re{SkewYoungTransferMatrix},
\be
t_{Y (\bit{\scriptstyle m}/\bit{\scriptstyle n})} (x)
=
\sum_{Y_\alpha} \prod_{\alpha (j,k) \in Y_\alpha}
p_{\alpha (j,k)} \mathcal{Y} \left( \alpha (j, k), u^{[\tilde{m}_1 - m_1 + 2j - 2k]} \right)
\, .
\ee
These transfer matrices admit a determinant representation in terms of (anti-)symmetric transfer
matrices $t_{[a]}$ and $t^{\{s\}}$,
\ba
\label{SkewYoungAllOrder}
t_{Y (\bit{\scriptstyle m}/\bit{\scriptstyle n})} (x)
\!\!\!&=&\!\!\!
\det_{1 \leq j, k \leq m_1}
t_{[\tilde{m}_j - \tilde{n}_k - j + k]}
\left( x^{[m_1 - \tilde{m}_1 + \tilde{m}_j + \tilde{n}_k - j - k + 1]} \right)
\nonumber\\
&=&\!\!\!
\det_{1 \leq j, k \leq \tilde{m}_1}
t^{\{m_k - n_j + j - k\}} \left( x^{[m_1 - \tilde{m}_1 - m_k - n_j + j + k - 1]} \right)
\, ,
\ea
with boundary conditions $t_{[a < 0]} = t^{\{s < 0\}} = 0$. For auxiliary space labelled by a
rectangular Young supertableaux, the all-order transfer matrices obey Hirota-type equations
\be
t_{[a]}^{\{s\}} (x^+) t_{[a]}^{\{s\}} (x^-)
=
t_{[a]}^{\{s + 1\}} (x) t_{[a]}^{\{s - 1\}} (x)
+
t_{[a + 1]}^{\{s\}} (x) t_{[a - 1]}^{\{s\}} (x)
\, .
\ee
These can be extended to arbitrary Young supertableaux upon proper choice of boundary conditions
as was recently discussed for short-range super-spin chains in Ref.\ \cite{KazSorZab07} generalizing
earlier considerations for classical Lie algebras \cite{LipWieZab97}.

\subsection{Baxter equations}

Finally, let us derive a set of closed equations for the polynomials \re{BaxterAllOrder}. Again
it suffices to consider the lowest-dimensional transfer matrices only. Using the generating
functions \re{GFantiAllOrder} and \re{GFsymmAllOrder}, we obtain explicit form of transfer
matrices in defining fundamental representation
\ba
\label{BaxterQ2}
t_{[1]} (x)
\!\!\!&=&\!\!\!
(x^-)^L
\frac{\widehat{Q}^{(3)} \left( u + \ft{i}{2}\right)}{\widehat{Q}^{(3)} \left( u - \ft{i}{2}\right)}
\left(
{\rm e}^{\Delta_- (x^-)} \frac{Q^{(2)} (u - i)}{Q^{(2)} (u)}
-
{\rm e}^{\frac{1}{2} \Delta_+ (x^-) + \frac{1}{2} \Delta_- (x^-)}
\right)
\nonumber\\
&+&\!\!\!
(x^+)^L
\frac{\widehat{Q}^{(1)} \left( u - \ft{i}{2}\right)}{\widehat{Q}^{(1)} \left( u + \ft{i}{2}\right)}
\left(
{\rm e}^{\Delta_+ (x^+)} \frac{Q^{(2)} (u + i)}{Q^{(2)} (u)}
-
{\rm e}^{\frac{1}{2} \Delta_+ (x^+) + \frac{1}{2} \Delta_- (x^+)}
\right)
\, ,
\ea
cf.\ Refs.\ \cite{BeiSta05,Bei06}, and the rest deferred to Appendix \ref{TransferAppendix}. Removing
excitation associated with either first $Q^{(1)} = 1$ or last $Q^{(3)} = 1$ node of the symmetric
Kac-Dynkin diagram in Fig.\ \ref{dynkin} or both, we reduce to all-order transfer matrices in either
$sl(2|1)$ \cite{Bel07} or $sl(2)$ \cite{Bel06} subsectors of the theory, respectively. Performing the
same steps as earlier in Sect.\ \ref{LOBaxterSymmetric}, we find that the Baxter equations take the
form of second order finite-difference equations with coefficients determined by either antisymmetric,
\ba
\label{BaxterQ1a}
t_{[2]} (x) \bar{t}_{[2]} (x)
\widehat{Q}^{(1)} (u)
\!\!\!&+&\!\!\!
x^L
{\rm e}^{\ft12 \Delta_- (x) + \ft12 \Delta_+ (x)}
t_{[2]} (x) \bar{t}_{[1]} (x^-)
\widehat{Q}^{(1)} (u + i)
\\
&+&\!\!\!
x^L
{\rm e}^{\ft12 \Delta_- (x) + \ft12 \Delta_+ (x)}
\bar{t}_{[2]} (x) t_{[1]} (x^+)
\widehat{Q}^{(1)} (u - i)
= 0
\, , \nonumber\\
\label{BaxterQ3a}
t_{[2]} (x) \bar{t}_{[2]} (x)
\widehat{Q}^{(3)} (u)
\!\!\!&+&\!\!\!
x^L
{\rm e}^{\ft12 \Delta_- (x) + \ft12 \Delta_+ (x)}
\bar{t}_{[2]} (x) t_{[1]} (x^-)
\widehat{Q}^{(3)} (u + i)
\\
&+&\!\!\!
x^L
{\rm e}^{\ft12 \Delta_- (x) + \ft12 \Delta_+ (x)}
t_{[2]} (x) \bar{t}_{[1]} (x^+)
\widehat{Q}^{(3)} (u - i)
= 0
\, , \nonumber
\ea
or symmetric transfer matrices and deformed by dressing factors
\ba
\label{BaxterQ1s}
&&
\left[ t^{\{2\}} (x) - t_{[1]} (x^+) t_{[1]} (x^-) \right]
\left[ \bar{t}^{\{2\}} (x) - \bar{t}_{[1]} (x^+) \bar{t}_{[1]} (x^-) \right]
\widehat{Q}^{(1)}(u)
\\
&&\qquad\quad \
- \,
x^L {\rm e}^{\ft12 \Delta_- (x) + \ft12 \Delta_+ (x)}
\bar{t}_{[1]} (x^-)
\left[ t^{\{2\}} (x) - t_{[1]} (x^+) t_{[1]} (x^-) \right]
\widehat{Q}^{(1)} (u + i)
\nonumber\\
&&\qquad\quad \
- \,
x^L {\rm e}^{\ft12 \Delta_- (x) + \ft12 \Delta_+ (x)}
t_{[1]} (x^+)
\left[ \bar{t}^{\{2\}} (x) - \bar{t}_{[1]} (x^+) \bar{t}_{[1]} (x^-) \right]
\widehat{Q}^{(1)} (u - i)
= 0
\, , \nonumber\\
\label{BaxterQ3s}
&&
\left[ t^{\{2\}} (x) - t_{[1]} (x^+) t_{[1]} (x^-) \right]
\left[ \bar{t}^{\{2\}} (x) - \bar{t}_{[1]} (x^+) \bar{t}_{[1]} (x^-) \right]
\widehat{Q}^{(3)} (u)
\\
&&\qquad\quad \
- \,
x^L {\rm e}^{\ft12 \Delta_- (x) + \ft12 \Delta_+ (x)}
t_{[1]} (x^-)
\left[ \bar{t}^{\{2\}} (x) - \bar{t}_{[1]} (x^+) \bar{t}_{[1]} (x^-) \right]
\widehat{Q}^{(3)} (u + i)
\nonumber\\
&&\qquad\quad \
- \,
x^L {\rm e}^{\ft12 \Delta_- (x) + \ft12 \Delta_+ (x)}
\bar{t}_{[1]} (x^+)
\left[ t^{\{2\}} (x) - t_{[1]} (x^+) t_{[1]} (x^-) \right]
\widehat{Q}^{(3)} (u - i)
= 0
\, . \nonumber
\ea
On top of these equations, the product of Baxter polynomials $Q^{(1,3)}$ obey consistency conditions
\ba
\label{Q1Q3RelationsAllOrder}
\widehat{Q}^{(1)} \left( u^{[+ a]} \right) \widehat{Q}^{(3)} \left( u^{[- a]} \right) t_{[a]} (x)
\!\!\!&=&\!\!\!
\widehat{Q}^{(1)} \left( u^{[- a]} \right) \widehat{Q}^{(3)} \left( u^{[+ a]} \right) \bar{t}_{[a]} (x)
\, , \\
\widehat{Q}^{(1)} \left( u^{[+ s]} \right) \widehat{Q}^{(3)} \left( u^{[- s]} \right) t^{\{s\}} (x)
\!\!\!&=&\!\!\!
\widehat{Q}^{(1)} \left( u^{[- s]} \right) \widehat{Q}^{(3)} \left( u^{[+ s]} \right) \bar{t}^{\{s\}} (x)
\, ,
\ea
with (anti-)symmetric transfer matrices, in their turn, obeying functional relations identical to the
one for a short-range magnet \re{TreeRelationsTransfer}. Once the nested Baxter polynomials are
determined from Eqs.\ \re{BaxterQ2} -- \re{BaxterQ3s}, they generate the spectrum of transfer matrices
associated with a skew Young supertableaux via Eq.\ \re{SkewYoungAllOrder}.

\section{Conclusions}

The main focus of the present study was a closed $psl(2|2)$ subsector of the dilatation operator
in maximally supersymmetric Yang-Mills theory. The sector is encoded into the $\mathcal{N} = 2$
Wess-Zumino supermultiplet embedded into the $\mathcal{N} = 4$ light-cone superfield and obeys
autonomous renormalization group evolution to all orders in 't Hooft coupling.

Due to non-uniqueness of simple root systems for superalgebras we chose the one which allows for
a straightforward generalization of Bethe equations to all orders of perturbation theory. We
concentrated on a Kac-Dynkin diagram having two isotropic odd nodes which reduces to noncompact
$sl(2)$ sector when the number of fermionic excitations vanishes. We used close relation of
transfer matrices to representation theory and analytic Bethe ansatz to construct their form
for auxiliary spaces associated with skew Young supertableaux in terms of Baxter functions. The
latter possess determinant representation in terms of transfer matrices with symmetric $(s)$
or antisymmetric $(1^a)$ atypical representations in the auxiliary space. For zero spectral
parameter $u = 0$ these relations reproduce well known supercharacter formulas for supergroups.
Bethe Ansatz equations ensure that these transfer matrices are pole-free at positions of nested
Bethe roots. We have used these equations in perturbative, asymptotic sense when the dynamical
poles at $u = g/(2 u_k)$ are not reachable. Proper incorporation of the latter into the formalism
and a proof of their cancellations would contribute to resolution of the notorious wrapping problem
for the underlying long-range spin chain, i.e., when the range of interaction is even or higher
them the length of chain itself.

We have formulated equivalent closed systems of Baxter equations \re{BaxterQ2}, \re{BaxterQ1a}
and \re{BaxterQ3a} or \re{BaxterQ3a}, \re{BaxterQ1s} and \re{BaxterQ3s} for nested Baxter functions.
Transfer matrices encode a full set of mutually commuting conserved quantities. The solutions to
these sets determine the spectra of quantized charges and roots of Baxter functions and thus
determine spectra of anomalous dimensions of Wilson operators to all orders in gauge coupling
constant via the equation
\be
\label{AllOrderAD}
\gamma (g) = i g^2 \int_{- 1}^1 \frac{dt}{\pi} \sqrt{1 - t^2}
\left(
\ln Q^{(2)} \left( \ft{i}{2} - g t \right)
-
\ln Q^{(2)} \left( - \ft{i}{2} - g t \right)
\right)^\prime
\, .
\ee
Perturbative solutions to these equations in lowest few orders in 't Hooft coupling were given
for the $sl(2|1)$ long-range spin chain in Ref.\ \cite{Bel07} and applies in a straightforward
fashion to the case at hand.

The framework of Baxter equations is advantageous for a number of reasons. While both Bethe
Ansatz and Baxter equations produce identical results for models based on representations with
highest and/or lowest weight vectors, the Baxter framework applies even when the pseudovacuum
state in the Hilbert space of the chain is absent. For noncompact super-spin chains, the number
of eigenstates is infinite for a finite length of the spin chain and the analysis of spectra
in this approach is preferable. Another advantage of Baxter approach within the present context
of the putative long-range magnet is that it clearly demonstrates the limitation of the asymptotic
equations, i.e., their invalidity beyond wrapping order. Since the Baxter equation is a
polynomial equation in the renormalized spectral parameter $x[u]$, by expanding everything in
perturbative series in coupling constant, one gets nonpolynomial terms. However, it is not this
non-polynomiality per se which breaks beyond wrapping order rather it is the fact that by solving
the system of equations stemming from coefficients in front of powers of $u$, one finds that it
becomes overdetermined and inconsistent. This may serve as a starting point to elaborate on corrections
terms which yield all-order anomalous dimensions even for short spin chains. Another problem which
deserves a dedicated study is the origin of the dressing phase. Recall that the eigenvalues of the
Baxter operator have a clear physical meaning of the wave functions of the magnet in separated
variables \cite{Skl88}, a property which should be preserved to all order in 't Hooft coupling. The
crossing symmetry \cite{Jan06,Bei05} implemented in terms of $Q$ yields a relations which should
be understood in physical terms as a certain constraint on the analytical properties of this wave
function. It remains to formulate $psu(2,2|4)$ Baxter equations for the full theory relying on the
analytic Bethe Ansatz, to start with.

\vspace{0.5cm}

This work was supported by the U.S. National Science Foundation under grant no. PHY-0456520.

\appendix

\section{Superconformal algebra and superspace realization}
\label{SuperspaceRealization}

The ${\cal N}$-extended superconformal algebra $su(2,2|\mathcal{N})$ contain 15 even charges
$\mathfrak{P}_\mu$, $\mathfrak{M}_{\mu\nu}$, $\mathfrak{D}$ and $\mathfrak{K}_\mu$ and $4 \mathcal{N}$
odd charges $\mathfrak{Q}_{\alpha A}$, $\bar{\mathfrak{Q}}^{\dot\alpha A}$, $\mathfrak{S}_\alpha^A$,
$\bar{\mathfrak{S}}^{\dot\alpha}_A$ which are two-component Weyl spinors carrying an $su(\mathcal{N})$
index $A = 1, 2, \dots, \mathcal{N}$. There are additional bosonic chiral charge $\mathfrak{R}$ and, in
case of extended $\mathcal{N} > 1$ supersymmetries, also $su(\mathcal{N})$ charges $\mathfrak{T}_A{}^B$
satisfying the commutation relations $[ \mathfrak{T}_A{}^B , \mathfrak{T}_C{}^D ] = \delta_C^B
\mathfrak{T}_A{}^D - \delta_A^D \mathfrak{T}_C{}^D $. The nontrivial commutation relations, on top of
conventional bosonic $su(2,2)$ relations, read
\ba
&&
\begin{array}{llllll}
{}\{ \mathfrak{Q}_{\alpha A} , \bar{\mathfrak{Q}}_{\dot\beta}^B \}
&\!\!\!
= 2 \delta_A^B \, \bar \sigma^\mu{}_{\alpha\dot\beta} \mathfrak{P}_\mu
\,, \qquad
&
{}[ \mathfrak{Q}_{\alpha A} , \mathfrak{M}^{\mu\nu} ]
&\!\!\!
= \frac12 \sigma^{\mu\nu}{}_\alpha{}^\beta \mathfrak{Q}_{\beta A}
\,, \qquad
&
{}[ \mathfrak{Q}_{\alpha A} , \mathfrak{D} ]
&\!\!\!
= \frac{i}2 \mathfrak{Q}_{\alpha A}
\,, \\ [1mm]
{}[ \mathfrak{Q}_{\alpha A} , \mathfrak{K}^\mu ]
&\!\!\!
= \bar\sigma^\mu{}_{\alpha\dot\beta} \bar{\mathfrak{S}}^{\dot\beta}_A
\,,
&
{}[ \mathfrak{S}_\alpha^A , \mathfrak{P}^\mu ]
&\!\!\!
= \bar\sigma^\mu{}_{\alpha \dot\beta} \bar{\mathfrak{Q}}^{\dot\beta A}
\,,
&
{}[ \mathfrak{S}_\alpha^A , \mathfrak{D} ]
&\!\!\!
= - \frac{i}2 \mathfrak{S}_\alpha^A
\,, \\ [1mm]
{}[ \mathfrak{S}_\alpha^A , \mathfrak{M}^{\mu\nu} ]
&\!\!\!
= \frac12 \sigma^{\mu\nu}{}_\alpha{}^\beta \mathfrak{S}_\beta^A
\,,
&
{}[ \mathfrak{Q}_{\alpha A} , \mathfrak{R} ]
&\!\!\!
= \mathfrak{Q}_{\alpha A}
\,,
&
{}[ \mathfrak{S}_\alpha^A , \mathfrak{R} ]
&\!\!\!
= - \mathfrak{S}_\alpha^A
\,, \\ [1mm]
{}[ \mathfrak{S}_\alpha^A , \bar{\mathfrak{S}}_{\dot\beta B} ]
&\!\!\!
= 2 \delta^A_B \, \bar\sigma^\mu{}_{\alpha\dot\beta} \mathfrak{K}_\mu
\,,
&
{}[ \mathfrak{T}_A{}^B, \mathfrak{Q}_{\alpha C} ]
&\!\!\!
= t_{AC}^{BD} {\cal Q}_{\alpha D}
\,,
&
{}[ \mathfrak{T}_A{}^B, \mathfrak{S}_\alpha^C ]
&\!\!\!
= - t_{AD}^{BC} \mathfrak{S}_\alpha^D
\,,
\end{array}
\nonumber\\ [1mm]
&&\qquad\qquad
\begin{array}{ll}
{}\{ \mathfrak{S}_\alpha^A , \mathfrak{Q}^\beta_B \}
&\!\!\!
= 2 i \delta^A_B \, \delta_\alpha^\beta \mathfrak{D}
+
\delta^A_B \, \sigma^{\mu\nu}{}_\alpha{}^\beta \mathfrak{M}_{\mu\nu}
+
\left( \frac{4}{{\cal N}} - 1 \right)
\delta^A_B \, \delta_\alpha^\beta \mathfrak{R}
-
4 \mathfrak{T}_B{}^A \delta_\alpha^\beta
\, ,
\end{array}
\ea
and their complex conjugate. Here $t^{AC}_{BD} = \delta^A_D \delta^C_B - \frac{1}{{\cal N}} \delta^A_B
\delta^C_D$. The remaining (anti-)commutators vanish. Here and below we use conventions for Clifford
algebra from Ref.\ \cite{BelDerKorMan03}.

The extended superspace with coordinates $\mathcal{X} = ( x^\mu , \theta^{\alpha A} ,
\bar\theta_{\dot\alpha A} )$ admits a coset manifold parameterization
\begin{equation}
\label{Coset}
g (\mathcal{X}) \equiv
{\rm e}^{
i x^\mu \mathfrak{P}_\mu
+
i \theta^{\alpha A} \mathfrak{Q}_{\alpha A}
+
i \bar\theta_{\dot\alpha A} \bar{\mathfrak{Q}}^{\dot\alpha A}
}
\, ,
\end{equation}
with the multiplication law
\be
g (\mathcal{X}_1) g (\mathcal{X}_2) = g (\mathcal{X}_3)
\, ,
\ee
and transformed coordinates being
\ba
\mathcal{X}
=
\left(
x_1^\mu + x_2^\mu -
i \theta_2^{\alpha A} \bar\sigma^\mu{}_{\alpha \dot\beta} \bar\theta_{1 A}^{\dot\beta}
+
i \theta_1^{\alpha A} \bar\sigma^\mu{}_{\alpha \dot\beta} \bar\theta_{2 A}^{\dot\beta}
\, ,
\theta_1^{\alpha A}
+
\theta_2^{\alpha A}
\, ,
\bar\theta_{1 \dot\alpha A}
+
\bar\theta_{2 \dot\alpha A}
\right)
\, .
\ea
In this parametrization, a superfield in this superspace is defined as
\begin{equation}
\Phi (\mathcal{X}) = g (\mathcal{X}) \Phi (0) g^{- 1} (\mathcal{X})
\, .
\end{equation}

Using conventional technique of induced representations one easily computes a representation
of generators in the superspace. The center elements are
\begin{eqnarray}
\begin{array}{ll}
{}[\mathfrak{M}_{\mu\nu} , \Phi(0)] = - \Sigma_{\mu\nu} \Phi (0)
\, ,
& \quad
{}[\mathfrak{D} , \Phi(0)] = - i d \Phi (0)
\, , \\
{}[\mathfrak{R} , \Phi(0)] = r \Phi (0)
\, ,
& \quad
{}[\mathfrak{T}_A{}^B , \Phi(0)] = t_A^B \Phi (0)
\, ,
\end{array}
\end{eqnarray}
all the rest vanish\footnote{We assumed here that the superfield $\Phi$ is a tensor with
respect to $SU(\mathcal{N})$ group. The superfields of gauge theories are $SU(\mathcal{N})$
scalars thus $t_A^B = 0$.}. The representation of generators as differential operators acting
on the coordinates $\mathcal{X}$ of the superfield $\Phi$,
\ba
\label{Rep1}
[ \mathfrak{G} , {\mit\Phi} (\mathcal{X}) ] \equiv G {\mit\Phi} (\mathcal{X})
\, ,
\ea
is for $su(2,2)$ bosonic
\ba
i P^\mu
\!\!\!&=&\!\!\!
\partial^\mu
\, , \\
i M^{\mu\nu}
\!\!\!&=&\!\!\!
x^\mu \partial^\nu
-
x^\nu \partial^\mu
-
i \Sigma^{\mu\nu}
-
\ft{i}{2} \theta^{\beta A} \,
\sigma^{\mu\nu}{\,}_\beta{}^\alpha
\partial_{\theta^{\alpha A}}
-
\ft{i}{2} \bar\theta_{\dot\beta A} \,
\bar\sigma^{\mu\nu}{\,}^{\dot\beta}{}_{\dot\alpha}
\partial_{\bar\theta_{\dot\alpha A}}
\, , \nonumber\\
i D
\!\!\!&=&\!\!\!
d
+
x^\mu \partial_\mu
+
\ft12
\theta^{\alpha A}
\partial_{\theta^{\alpha A}}
+
\ft12
\bar\theta_{\dot\alpha A}
\partial_{\bar\theta_{\dot\alpha A}}
\, , \nonumber\\
i K^\mu
\!\!\!&=&\!\!\!
\left(
2 x^\mu x^\nu - x^2 g^{\mu \nu}
\right)
\partial_\nu
+
2
\bar\theta_{\dot\alpha A} \sigma^{\mu \, \dot\alpha \beta} \theta^B_{\beta}
\,
\bar\theta_{\dot\beta B} \sigma^{\nu \, \dot\beta \alpha} \theta_\alpha^A
\partial_\nu
+
2 d x^\mu
-
2 i x_\nu \Sigma^{\mu\nu}
\nonumber\\
&+&\!\!\!
i
\varepsilon^{\mu\nu\rho\sigma}
\bar\theta_{\dot\alpha A} \sigma_\nu{}^{\dot\alpha \beta} \theta_\beta^A
\Sigma_{\rho\sigma}
+
i r \left( \ft{4}{\mathcal{N}} - 1 \right)
\bar\theta_{\dot\alpha A} \sigma^{\mu \, \dot\alpha \beta} \theta_\beta^A
-
4 i
\bar\theta_{\dot\alpha A} \sigma^{\mu \, \dot\alpha \beta} \theta_\beta^B
t_B^A
\nonumber\\
&+&\!\!\!
(
x_\nu \theta^{\alpha A} \bar\sigma^\mu{}_{\alpha\dot\beta}
\sigma^\nu{}^{\dot\beta \gamma}
-
2 i
\bar\theta_{\dot\alpha B} \sigma^{\mu \, \dot\alpha \beta} \theta_\beta^A
\,
\theta^{\gamma B}
)
\partial_{\theta^{\gamma A}}
+
(
x_\nu \bar\theta_{\dot\alpha A} \sigma^{\mu \, \dot\alpha\beta}
\bar\sigma^\nu{}_{\beta \dot\gamma}
+
2 i
\bar\theta_{\dot\alpha A} \sigma^{\mu \, \dot\alpha \beta} \theta_\beta^B
\,
\bar\theta_{\dot\gamma B}
)
\partial_{\bar\theta_{\dot\gamma A}}
\, , \nonumber
\ea
and fermionic generators
\ba
i Q_{\alpha A}
\!\!\!&=&\!\!\!
\partial_{\theta^{\alpha A}}
+
i \bar\sigma^\mu{\,}_{\alpha \dot\beta} \, \bar\theta^{\dot\beta}_A
\partial_\mu
\, , \\
i \bar{Q}{}^{\dot\alpha A}
\!\!\!&=&\!\!\!
\partial_{\bar\theta_{\dot\alpha A}}
+
i \sigma^\mu{\,}^{\dot\alpha \beta} \, \theta_\beta^A
\partial_\mu
\, , \nonumber\\
i S_\alpha^A
\!\!\!&=&\!\!\!
- \left( 2 d + r \left( \ft{4}{\mathcal{N}} - 1 \right) \right)
\theta_\alpha^A
+
\sigma^{\mu\nu}{}_\alpha{}^\beta \theta_\beta^A \Sigma_{\mu\nu}
+
4 t_B^A \theta_\alpha^B
\nonumber\\
&-&\!\!\!
(
x_\nu \bar\sigma^\nu{}_{\alpha \dot\beta} \sigma^{\mu \, \dot\beta\gamma}
\theta_\gamma^A
-
2 i
\theta_\alpha^B
\,
\bar\theta_{\dot\alpha B} \sigma^{\mu \, \dot\alpha \beta} \theta_\beta^A
)
\partial_\mu
+
4 \theta_\alpha^B \theta^{\beta A}
\partial_{\theta^{\beta B}}
-
\left(
2 \theta_\alpha^B \bar\theta_{\dot\beta B}
-
i x_\mu \bar\sigma^\mu{}_{\alpha\dot\beta}
\right)
\partial_{\bar\theta_{\dot\beta A}}
\, , \nonumber\\
i \bar{S}{}^{\dot\alpha}_A
\!\!\!&=&\!\!\!
- \left( 2 d - r \left( \ft{4}{\mathcal{N}} - 1 \right) \right)
\bar\theta^{\dot\alpha}_A
+
\bar\sigma^{\mu\nu}{\,}^{\dot\alpha}{}_{\dot\beta}
\bar\theta^{\dot\beta}_A
\Sigma_{\mu\nu}
-
4 t_A^B \bar\theta^{\dot\alpha}_B
\nonumber\\
&-&\!\!\!
(
x_\nu \sigma^{\nu \, \dot\alpha \beta} \bar\sigma^\mu{}_{\beta\dot\gamma}
\bar\theta^{\dot\gamma}_A
+
2 i
\bar \theta^{\dot\alpha}_B
\,
\bar\theta_{\dot\beta A} \sigma^{\mu \, \dot\beta \alpha} \theta_\alpha^B
)
\partial_\mu
+
4 \bar\theta^{\dot\alpha}_B \bar\theta_{\dot\beta A}
\partial_{\bar\theta_{\dot\beta B}}
-
\left(
2 \bar\theta^{\dot\alpha}_B \theta^{\beta B}
-
i x_\mu \sigma^{\mu \, \dot\alpha\beta}
\right)
\partial_{\theta^{\beta A}}
\, , \nonumber
\ea
and finally $su(\mathcal{N})$ and $u(1)$ generators
\ba
i T_B{}^A
\!\!\!&=&\!\!\!
i t_B^A
+
i
\left(
\theta^{\alpha A} \partial_{\theta^{\alpha B}}
-
\bar\theta_{\dot\alpha B} \partial_{\bar\theta_{\dot\alpha A}}
\right)
-
\ft{i}{{\cal N}}
\delta_B^A
\left(
\theta^{\alpha C} \partial_{\theta^{\alpha C}}
-
\bar\theta_{\dot\alpha C}
\partial_{\bar\theta_{\dot\alpha C}}
\right)
\, , \\
i R
\!\!\!&=&\!\!\!
i r
- i \theta^{\alpha A} \partial_{\theta^{\alpha A}}
+ i \bar\theta_{\dot\alpha A}
\partial_{\bar\theta_{\dot\alpha A}}
\, , \nonumber
\ea
cf.\ Ref.\ \cite{GatGriRocSie83}.

\subsection{Light-cone reduction of superspace}

In this paper we are interested in a closed $sl(2|2)$ subsector of the dilatation operator, which
acts on the light-cone composite operators \re{LCoperator} built from $\mathcal{N} = 2$ Wess-Zumino
superfield of the truncated maximally supersymmetric gauge theory. The $sl(2|2)$ subalgebra of the
superconformal algebra is spanned by the following generators
\be
\mathfrak{P}^+
\, , \quad
\mathfrak{M}^{-+}
\, , \quad
\mathfrak{D}
\, , \quad
\mathfrak{K}^-
\, , \quad
\mathfrak{Q}_{+ \alpha A}
\, , \quad
\bar{\mathfrak{Q}}_{+}^{\dot\alpha A}
\, , \quad
\mathfrak{S}_{- \alpha}^A
\, , \quad
\bar{\mathfrak{S}}_{- A}^{\dot\alpha}
\, , \quad
\mathfrak{R}
\, , \quad
\mathfrak{T}_A{}^B
\, , \quad
\mathfrak{M}^{12}
\, .
\ee
The vector indices of generators are contracted with the light-cone vectors $n^\mu$ and $n^{\ast\mu}$
obeying the following conditions $n^2 = n^{\ast 2} = 0$ and $n \cdot n^\ast = 1$, such that
\be
\mathfrak{G}^\mu n_\mu = \mathfrak{G}^+
\, , \qquad
\mathfrak{G}^\mu n^\ast_\mu = \mathfrak{G}^-
\, .
\ee
While the fermionic generators are obtained with the help of the projectors
\be
\label{LCprojection}
\mathfrak{G}_{\pm \alpha}
=
\ft12 \bar\sigma^\mp{}_{\alpha\dot\beta} \, \sigma^{\pm \; \dot\beta\gamma} \mathfrak{G}_\gamma
\, , \qquad
\bar{\mathfrak{G}}_\pm^{\dot\alpha}
=
\ft12 \sigma^{\mp \; \dot\alpha\beta} \, \bar\sigma^\pm{}_{\beta\dot\gamma}
\bar{\mathfrak{G}}^{\dot\gamma}
\, .
\ee
One can easily convince oneself that actually only one component in each light-cone Weyl spinor is
nonvanishing. This reflect a general phenomenon of the light-cone formalism: a spinor satisfying
the Weyl condition can be described by a complex Grassmann variable without a Lorentz index. Further,
it is convenient to introduce the following combinations of the generators
\be
\label{Redefinitions}
\begin{array}{llll}
i \mathfrak{P}^+ \equiv - \mathfrak{L}^-
\, , \ \
&
\ft{i}{2} \mathfrak{K}^- \equiv \mathfrak{L}^+
\, , \ \
&
\ft{i}{2} ( \mathfrak{D} + \mathfrak{M}^{-+} )
\equiv \mathfrak{L}^0
\, , \ \
&
\ft14 (\ft{4}{{\cal N}} - 1) \mathfrak{R} - \ft12 \mathfrak{M}^{12}
\equiv \mathfrak{B}
\, , \\ [1mm]
i \mathfrak{Q}_{1 A} \equiv \sqrt[4]{8} \, \mathfrak{V}^-_A
\, , \ \
&
i \bar{\mathfrak{Q}}{}_{\dot 1}^A \equiv - i \sqrt[4]{8} \, \bar{\mathfrak{V}}^{A, -}
\, , \ \
&
i \mathfrak{S}^{1 A} \equiv - \sqrt[4]{32} \,  \bar{\mathfrak{V}}^{A, +}
\, , \ \
&
i \bar{\mathfrak{S}}{}^{\dot 1}_A \equiv i \sqrt[4]{32} \,  \mathfrak{V}^+_A
\, .
\end{array}
\ee
The representation \re{Rep1} of these generators in the light-cone superspace then reads for
the bosonic $sl(2)$ subalgebra
\ba
\label{LCsl2}
L^- \!\!\!&=&\!\!\! - \partial_z
\, , \\
L^+ \!\!\!&=&\!\!\!
2 \ell z
+
\bar\theta_A \theta^B \left( b \, \delta^A_B - t_B^A \right)
+
\left( z^2 + \ft14 ( \bar\theta_A \theta^A )^2 \right) \partial_z
+
\left( z + \ft12 \bar\theta_B \theta^B \right)
\theta^A \partial_{\theta^A}
+
\left( z - \ft12 \bar\theta_B \theta^B \right)
\bar\theta_A \partial_{\bar\theta_A}
\, , \nonumber\\
L^0
\!\!\!&=&\!\!\!
\ell + z \partial_z
+ \ft12 \theta^A \partial_{\theta^A}
+ \ft12 \bar\theta_A \partial_{\bar\theta_A}
\, . \nonumber
\ea
The fermionic generators read
\ba
\label{LCfer}
V^-_A
\!\!\!&=&\!\!\!
\partial_{\theta^A} + \ft12 \, \bar\theta_A \, \partial_z
\, , \\
\bar{V}^{A,-}
\!\!\!&=&\!\!\!
\partial_{\bar\theta_A}
+
\ft12 \, \theta^A \, \partial_z
\, , \nonumber\\
V^+_A
\!\!\!&=&\!\!\!
\left( (\ell - b) \delta_A^B + t_A^B \right) \bar\theta_B
+
\ft12
\left( z - \ft12 \bar\theta_B \theta^B \right) \bar\theta_A \partial_z
+
\bar\theta_A \bar\theta_B \partial_{\bar\theta_B}
+
\left( z + \ft12 \bar\theta_B \theta^B \right) \partial_{\theta^A}
\, , \nonumber\\
\bar{V}^{A,+}
\!\!\!&=&\!\!\!
\left( (\ell + b) \delta^A_B - t_B^A \right) \theta^B
+
\ft12 \left( z + \ft12 \bar\theta_B \theta^B \right) \theta^A \partial_z
+
\theta^A \theta^B \partial_{\theta^B}
+
\left( z - \ft12 \bar\theta_B \theta^B \right) \partial_{\bar\theta_A}
\, , \nonumber
\ea
and the remaining are
\ba
\label{LCsun}
B
\!\!\!&=&\!\!\!
b + \ft12 \left( 1 - \ft{2}{{\cal N}} \right)
\left(
\theta^A \partial_{\theta^A} - \bar\theta_A \partial_{\bar\theta_A}
\right)
\, , \\
T_B{}^A
\!\!\!&=&\!\!\!
t_B^A
+
\left( \theta^A \partial_{\theta^B} - \bar\theta_B \partial_{\bar\theta_A}
\right)
-
\ft{1}{\mathcal{N}} \delta_B^A
\left( \theta^C \partial_{\theta^C} - \bar\theta_C \partial_{\bar\theta_C} \right)
\, . \nonumber
\ea
Here we used a notation $z = x^-$ for the coordinate projected on the light cone. The Grassmann
coordinates $\theta^A$ and $\bar\theta_A$ were redefined compared to the ones in the covariant
superspace \re{Coset} as follows
\be
\label{RedefLCweyl}
\theta^A \equiv \sqrt[4]{8} \, \theta^{1 A}
\, , \qquad
\bar\theta_A \equiv i \sqrt[4]{8} \, \bar\theta^{\dot 1}_A
\, .
\ee
The quantum numbers of the superfields are encoded into the conformal spin $\ell$ and chirality $b$
\be
\ell = \ft{1}{2} (s + d)
\, , \qquad
b = \ft{1}{4} r \left( \ft{4}{\mathcal{N}} - 1\right) - \ft{1}{2} h
\, ,
\ee
with the latter being a linear combination of its helicity $h = - \Sigma^{12}$ and the
$R-$charge $r$.

The commutation relations between the generators in the representation \re{LCsl2}, \re{LCfer} and
\re{LCsun} can be summarized by combining them in an $sl(2|\mathcal{N})-$covariant matrix with
components
\be
\label{sl2Ngenerators}
\begin{array}{lll}
E^{00} = L^0 - \ft{\mathcal{N}}{\mathcal{N} - 2} B
\, , \
&
E^{0A} = - V^+_A
\, , \
&
E^{0, \mathcal{N} + 1} = L^+
\, , \\ [3mm]
E^{A0} = - \bar{V}^{A, -}
\, , \ &
E^{AB}
= T_B{}^A + \ft{2}{\mathcal{N} - 2} B \delta_B^A
\, , \ &
E^{A, \mathcal{N} + 1} = - \bar{V}^{A, +}
\, , \\ [3mm]
E^{\mathcal{N} + 1, 0} = L^-
\, , \
&
E^{\mathcal{N} + 1, A} = V^-_A
&
E^{\mathcal{N} + 1, \mathcal{N} + 1}
=
- L^0 - \ft{\mathcal{N}}{\mathcal{N} - 2} B
\, .
\end{array}
\ee
Then the graded commutation relations read in a concise form
\ba
{}[E^{\mathcal{AB}} , E^{\mathcal{CD}}\}
\!\!\!&\equiv&\!\!\!
E^{\mathcal{AB}} E^{\mathcal{CD}}
-
(-1)^{(\bar{\mathcal{A}} + \bar{\mathcal{B}})(\bar{\mathcal{C}} + \bar{\mathcal{D}})}
E^{\mathcal{CD}} E^{\mathcal{AB}}
\nonumber\\
&=&\!\!\!
\delta^{\mathcal{CB}} E^{\mathcal{AD}}
-
(-1)^{(\bar{\mathcal{A}} + \bar{\mathcal{B}})(\bar{\mathcal{C}} + \bar{\mathcal{D}})}
\delta^{\mathcal{AD}} E^{\mathcal{CB}}
\, ,
\ea
where the indices $\mathcal{A}, \dots$ run over $\mathcal{N} + 2$ values $(0, A, \mathcal{N} + 1)$
and possess the gradings $\bar{0} = \overline{\mathcal{N} + 1} = 0$ and $\bar A = 1$.

\subsection{Projection to $psl(2|2)$ subsector}

The superfield \re{N4-superfield} encoding the field content of maximally supersymmetric Yang-Mills
theory is chiral, thus the dependence on the Grassmann variables $\bar\theta^A$ is trivial, i.e.,
$\Phi (z, \theta^a, \bar\theta_A) = \Phi (z + \ft12 \bar\theta_A \theta^A, \theta^A, 0)$. This
simplifies significantly the form of the generators, which are quoted in Ref.\ \cite{BelDerKorMan04}.
The generators on the $sl(2|2)$ subsector of $\mathcal{N} = 4$ theory are found making use of the
superfield truncation \re{TruncationN4toN2}, such that
\be
E \Phi (z, \theta^A)|_{\theta^1 = 0}
=
{\dots} + \theta^4 \mathcal{E} \Psi_{\scriptscriptstyle\rm WZ} (\mathcal{Z})
\, ,
\ee
where $E$ are given in Eqs.\ \re{sl2Ngenerators} for $\mathcal{N} = 4$ and $\mathcal{E}$ are
generators in question acting on a function of $Z = (z, \vartheta^a)$. One finds the following
identification of $sl(2|2)$ generators $\mathcal{E}^{\alpha\beta}$, obeying the commutation
relations
\ba
\label{sl22}
{}[\mathcal{E}^{\alpha \beta} , \mathcal{E}^{\gamma \delta}\}
\!\!\!&\equiv&\!\!\!
\mathcal{E}^{\alpha \beta} \mathcal{E}^{\gamma \delta}
-
(-1)^{(\bar{\alpha} + \bar{\beta})(\bar{\gamma} + \bar{\delta})}
\mathcal{E}^{\gamma \delta} \mathcal{E}^{\alpha \beta}
\nonumber\\
&=&\!\!\!
\delta^{\gamma \beta} \mathcal{E}^{\alpha \delta}
-
(-1)^{(\bar{\alpha} + \bar{\beta})(\bar{\gamma} + \bar{\delta})}
\delta^{\alpha \delta} \mathcal{E}^{\gamma \beta}
\, ,
\ea
to the ones of the truncated $\mathcal{N} = 2$ superconformal light-cone algebra
\be
\label{sl22generators}
\begin{array}{lll}
\mathcal{E}^{00} = \mathcal{L}^0 - 2 \mathcal{B}
\, , \quad
&
\mathcal{E}^{0a} = - \mathcal{V}^+_a
\, , \quad
&
\mathcal{E}^{03} = \mathcal{L}^+
\, , \\ [3mm]
\mathcal{E}^{a0} = - \bar{\mathcal{V}}^{a, -}
\, , \quad
&
\mathcal{E}^{ab}
= \mathcal{T}_b{}^a + 2 \mathcal{B} \delta_b^a
\, , \quad
&
\mathcal{E}^{a3} = - \bar{\mathcal{V}}^{a, +}
\, , \\ [3mm]
\mathcal{E}^{30} = \mathcal{L}^-
\, , \quad
&
\mathcal{E}^{3a} = \mathcal{V}^-_a
\, , \quad
&
\mathcal{E}^{33}
=
- \mathcal{L}^0 - 2 \mathcal{B}
\, .
\end{array}
\ee
As it is obvious from this formula, the indices of $\mathcal{E}^{\alpha\beta}$ run over $\alpha = 0,
a, 3$ with an $su(2)$ index taking two values $a = 1,2$ and the grading reflects the one of
\re{sl2Ngenerators}, i.e., $\bar{0} = \bar{3} = 0$ and $\bar{a} = 1$. The above generators are
realized on $\Psi_{\scriptscriptstyle\rm WZ}$ superfield \re{N2WZsuperfield} and read
\be
\label{psl22}
\begin{array}{llll}
\mathcal{L}^- = -\partial_z
\, , \
&
\mathcal{L}^+ =
z^2\partial_z + z \vartheta^a \partial_{\vartheta^a}
\, , \
&
\mathcal{L}^0 =
z \partial_z + \ft12 \vartheta^a \partial_{\vartheta^a}
\, , \
&
\mathcal{T}_b{}^a
=
\vartheta^a \partial_{\vartheta^b} - \ft12 \delta_b^a \vartheta^c \partial_{\vartheta^c}
\, , \\ [3mm]
\bar{\mathcal{V}}^{a,-}
=
\vartheta^a \partial_z \, , \
&
\bar{\mathcal{V}}^{a,+}
=
\vartheta^a (z \partial_z + \vartheta^c \partial_{\vartheta^c})
\, , \
&
\mathcal{V}^-_{a} = \partial_{\vartheta^a}
\, , \
&
\mathcal{V}^+_{a} = z \partial_{\vartheta^a} \, .
\end{array}
\ee
The $u_B(1)$ outer automorphism $\mathcal{B}$,
\be
\mathcal{B} = - \ft14 + \ft14 \vartheta^a \partial_{\vartheta^a}
\, ,
\ee
does not enter the right-hand side of the commutation relation of the generators \re{psl22}. Thus
the $sl(2|2)$ algebra \re{sl22} of generators \re{sl22generators} is a semidirect product $u_B(1)
\ltimes psu(2|2)$ with generators \re{psl22} forming the projective algebra $psl(2|2)$. Its quadratic
Casimir operator is
\ba
\mathbb{C}_2
\!\!\!&=&\!\!\!
\ft12 \sum_{\alpha,\beta = 0}^3 (-1)^{\bar\beta}
\mathcal{E}^{\alpha\beta} \mathcal{E}^{\beta\alpha}
\\
&=&\!\!\!
(\mathcal{L}^0 + \mathcal{T}^0)(\mathcal{L}^0 - \mathcal{T}^0 + 1)
+
\mathcal{L}^+ \mathcal{L}^- + \mathcal{T}_2{}^1 \mathcal{T}_1{}^2
-
\mathcal{V}^+_a \bar{\mathcal{V}}^{a,-}
-
\bar{\mathcal{V}}^{a,+} \mathcal{V}^-_a
\, , \nonumber
\ea
where we introduced a notation for $\mathcal{T}^0 \equiv \mathcal{T}_1{}^1 = - \mathcal{T}_2{}^2$.
Notice that this $su(2)$ generator is related to the one of $su(4)$ internal rotations as
$\mathcal{T}^0 = T_2{}^2 - T_3{}^3$.

\section{Serre-Chevalley bases for $psl(2|2)$}
\label{SerreChevalleyAppendix}

The projective algebra $psl (2|2)$ has rank two, but it is described by Kac-Dynkin diagrams having
three nodes. This implies that the simple roots of the root system are not linearly independent.
The root system is expressed in terms of weights\footnote{Our ordering of basis vectors reflects
the grading of the matrix of $sl(2|2)$ generators $\mathcal{E}^{\alpha \beta}$.} $\bit{v}_\alpha
= (\bit{\varepsilon}_0 | \bit{\delta}_1, \bit{\delta}_2| \bit{\varepsilon}_3)$ which form a basis
in the dual space of the Cartan subalgebra. The weights obey the conditions
\be
\label{LinearDepCartan}
\bit{\varepsilon}_0 + \bit{\varepsilon}_3 = 0
\, , \qquad
\bit{\delta}_1 + \bit{\delta}_2 = 0
\, .
\ee
and are endowed with a bilinear form such that
\be
(\bit{\varepsilon}_0 | \bit{\varepsilon}_0) = 1
\, , \qquad
(\bit{\delta}_1 | \bit{\delta}_1) = - 1
\, , \qquad
(\bit{\varepsilon}_0 | \bit{\delta}_1) = 0
\, .
\ee
The set of nonzero roots $\Delta = \Delta_0 \cup \Delta_1$ is divided into the set of even
$\Delta_0$ and odd $\Delta_1$ roots,
\be
\Delta_0 = \{ \bit{\varepsilon}_a - \bit{\varepsilon}_b , \bit{\delta}_a - \bit{\delta}_b \}
\, , \qquad
\Delta_1 = \{ \bit{\varepsilon}_a - \bit{\delta}_b , \bit{\delta}_b - \bit{\varepsilon}_a \}
\, .
\ee
The root vectors associated to these roots are
\be
\mathcal{E}^{\alpha \beta} \, \leftrightarrow \, \bit{v}_\alpha - \bit{v}_\beta
\, ,
\ee
and the Cartan subalgebra is spanned by the elements
\be
\mathcal{E}^{00} + \mathcal{E}^{11}
\, , \qquad
\mathcal{E}^{11} - \mathcal{E}^{22}
\, , \qquad
\mathcal{E}^{22} + \mathcal{E}^{33}
\, .
\ee
It is obvious from explicit realization that these generators are linearly dependent, exhibiting
peculiarities of projective algebras. There are several choices of simple root systems depending
on choices of Borel subalgebras. Let us discuss two simple root systems used in the main text,
i.e., corresponding to the distinguished and symmetric Kac-Dynkin diagram with two isotropic fermionic
roots.

\subsection{Distinguished Kac-Dynkin diagram}

The distinguished Kac-Dynkin diagram corresponding to the BBFF grading\footnote{Note that the
Kac-Dynkin diagrams for superalgebras are ambiguous. There is yet another distinguished diagram
with FFBB grading.} possesses the following ordering of the basis elements of the dual Cartan
subalgebra $( \bit{\varepsilon}_0, \bit{\varepsilon}_3 | \bit{\delta}_1, \bit{\delta}_2 )$ and
yields the simple root system\footnote{The normalization factors reflect linear dependence
\re{LinearDepCartan} of basis elements and are introduced in order to have conventional
definition of the Cartan matrix $A_{pq} = (\bit{\alpha}_p|\bit{\alpha}_q)$.}
\be
\bit{\alpha}_1 = \ft{1}{\sqrt{2}} ( \bit{\varepsilon}_0 - \bit{\varepsilon}_3 )
\, , \qquad
\bit{\alpha}_2 = \ft{1}{\sqrt{2}} ( \bit{\varepsilon}_3 - \bit{\delta}_1)
\, , \qquad
\bit{\alpha}_3 = \ft{1}{\sqrt{2}} ( \bit{\delta}_1 - \bit{\delta}_2 )
\, .
\ee
The Cartan subalgebra is formed by the generators
\be
h_1 = \mathcal{E}^{00} - \mathcal{E}^{33} = 2 \mathcal{L}^0
\, , \qquad
h_2 = \mathcal{E}^{33} + \mathcal{E}^{11} = \mathcal{T}^0 - \mathcal{L}^0
\, , \qquad
h_3 = - \mathcal{E}^{11} + \mathcal{E}^{22} = - 2 \mathcal{T}^0
\, .
\ee
and together with positive and negative root vectors
\be
\begin{array}{lll}
e^+_1 = \mathcal{E}^{03} = \mathcal{L}^+
\, , \qquad
&
e^+_2 = \mathcal{E}^{31} = \mathcal{V}^-_1
\, , \qquad
&
e^+_3 = \mathcal{E}^{12} = \mathcal{T}_2{}^1
\, , \\
e^-_1 = \mathcal{E}^{30} = \mathcal{L}^-
\, , \qquad
&
e^-_2 = \mathcal{E}^{13} = - \bar{\mathcal{V}}^{1,+}
\, , \qquad
&
e^-_3 = \mathcal{E}^{21} = \mathcal{T}_1{}^2
\, ,
\end{array}
\ee
form the Serre-Chevalley basis obeying the algebra
\be
\begin{array}{lll}
{}[h_p, e^\pm_q] = \pm A_{pq} e^\pm_q
\, , \qquad
&
{}[h_p, h_q] = 0
\, , \qquad
&
{}[e^\pm_p, e^\mp_q]_{p \neq q} = 0
\, , \\
{}[e_1^+, e_1^-] = h_1
\, , \qquad
&
{}\{e_2^+, e_2^-\} = h_2
\, , \qquad
&
{}[e_3^+, e_3^-] = - h_3
\, ,
\end{array}
\ee
with the Cartan matrix given in Eq.\ \re{CartanDistinguished}. From these one can generate
the entire algebra. The linear dependence of Cartan generators results in linear dependence
of the roots
\be
\bit{\alpha}_1 + 2 \bit{\alpha}_2 + \bit{\alpha}_3 = 0
\, .
\ee

\subsection{Symmetric Kac-Dynkin diagram}

The symmetric Kac-Dynkin diagram with two isotropic fermionic roots corresponding to FBBF grading
with the dual basis $(\bit{\delta}_1 | \bit{\varepsilon}_0, \bit{\varepsilon}_3 | \bit{\delta}_2)$
possesses the simple root system
\be
\bit{\alpha}_1 = \ft{1}{\sqrt{2}} ( \bit{\delta}_1 - \bit{\varepsilon}_0 )
\, , \qquad
\bit{\alpha}_2 = \ft{1}{\sqrt{2}} ( \bit{\varepsilon}_0 - \bit{\varepsilon}_3 )
\, , \qquad
\bit{\alpha}_3 = \ft{1}{\sqrt{2}} ( \bit{\varepsilon}_3 - \bit{\delta}_2 )
\, .
\ee
The Serre-Chevalley basis is
\be
\!\!\!
\begin{array}{lll}
h_1 = - \mathcal{E}^{11} - \mathcal{E}^{00} = - \mathcal{L}^0 - \mathcal{T}^0
\, , \qquad
&
h_2 = \mathcal{E}^{00} - \mathcal{E}^{33} = 2 \mathcal{L}^0
\, , \qquad
&
h_3 = \mathcal{E}^{33} + \mathcal{E}^{22} = - \mathcal{L}^0 - \mathcal{T}^0
, \\
e^+_1 = \mathcal{E}^{10} = - \bar{\mathcal{V}}^{1, -}
\, , \qquad
&
e^+_2 = \mathcal{E}^{03} = \mathcal{L}^+
\, , \qquad
&
e^+_3 = \mathcal{E}^{32} = \mathcal{V}_2^-
\, , \\
e^-_1 = \mathcal{E}^{01} = - \mathcal{V}^+_1
\, , \qquad
&
e^-_2 = \mathcal{E}^{30} = \mathcal{L}^-
\, , \qquad
&
e^-_3 = \mathcal{E}^{23} = - \bar{\mathcal{V}}^{2, +}
\, ,
\end{array}
\ee
with their commutation relations determined by the Cartan matrix \re{CartanSymmetric},
\be
\begin{array}{lll}
{}[h_p, e^\pm_q] = \pm A_{pq} e^\pm_q
\, , \qquad
&
{}[h_p, h_q] = 0
\, , \qquad
&
{}[e^\pm_p, e^\mp_q \}_{p \neq q} = 0
\, , \\
{}\{ e_1^+, e_1^- \} = - h_1
\, , \qquad
&
{}[e_2^+, e_2^-] = h_2
\, , \qquad
&
{}\{ e_3^+, e_3^- \} = h_3
\, .
\end{array}
\ee

\section{Excitation numbers}
\label{AppendixExcitation}

Here we will present the oscillator realization of $sl(2|2)$ algebra which is useful in relating
the number of excitations in nested Bethe Absatz to the eigenvalues of Cartan generators
\cite{FoeKar92}. One introduces \cite{GunSac82,FraSorSci96,BeiSta05,Bei04} bosonic and fermionic
raising $(\mathfrak{a}^\dagger, \mathfrak{b}^\dagger, \mathfrak{c}^{a \dagger})$ and lowering
$(\mathfrak{a}, \mathfrak{b}, \mathfrak{c}_a)$ operators, which obey the commutation relations
\be
{}[\mathfrak{a}, \mathfrak{a}^\dagger] = 1
\, , \qquad
{}[\mathfrak{b}, \mathfrak{b}^\dagger] = 1
\, , \qquad
\{\mathfrak{c}_a, \mathfrak{c}^{b \dagger}\} = \delta_a^b
\, .
\ee
The generators then read\footnote{These have hermitian conjugation properties identical to ones
in differential representation \re{psl22} endowed with $sl(2|2)$ invariant scalar product
\re{sl22ScalarProduct}, see Ref.\ \cite{BelDerKorMan05}.}
\ba
&&
\begin{array}{llll}
\mathcal{L}^- = - \mathfrak{a} \mathfrak{b}
\, , \
&
\mathcal{L}^+ =
\mathfrak{a}^\dagger \mathfrak{b}^\dagger
\, , \
&
\mathcal{L}^0 =
\ft12
\mathfrak{a}^\dagger
\mathfrak{a}
+
\ft12
\mathfrak{b}^\dagger
\mathfrak{b}
+
\ft12
\, , \
&
\mathcal{T}_b{}^a
=
\mathfrak{c}^{a \dagger} \mathfrak{c}_a
-
\ft12 \delta^a_b
\mathfrak{c}^{c \dagger} \mathfrak{c}_c
\, , \\ [3mm]
\bar{\mathcal{V}}^{a,-}
=
\mathfrak{a} \mathfrak{c}^{a \dagger}
\, , \
&
\bar{\mathcal{V}}^{a,+}
=
\mathfrak{b}^\dagger \mathfrak{c}^{a \dagger}
\, , \
&
\mathcal{V}^-_{a}
=
\mathfrak{b} \mathfrak{c}_a
\, , \
&
\mathcal{V}^+_{a}
=
\mathfrak{a}^\dagger \mathfrak{c}_a
\, ,
\end{array}
\nonumber\\ [2mm]
&&
\qquad\qquad\qquad\qquad\qquad\qquad
\mathcal{B}
=
- \ft14 \mathfrak{a}^\dagger \mathfrak{a} + \ft14 \mathfrak{b}^\dagger \mathfrak{b}
\, ,
\ea
and the oscillators obey the vanishing central charge condition
\be
\mathfrak{a}^\dagger \mathfrak{a} - \mathfrak{b}^\dagger \mathfrak{b}
+ \mathfrak{c}^{a \dagger} \mathfrak{c}_a - 1 = 0
\, .
\ee

The nested Bethe Ansatz for symmetric Kac-Dynkin diagram is built on an $L-$site first level
pseudovacuum state
\be
| \Omega \rangle_L = | Z_1 Z_2 \dots Z_L \rangle
\, ,
\ee
which in the basis of local Wilson operators corresponds to the product $Z^L (0)$. There are four
different types of excitations on each spin chain site\footnote{This state is created from true
vacuum with $su(4)$ oscillators $(\mathfrak{d}_1, \mathfrak{c}_1, \mathfrak{c}_2, \mathfrak{d}_2)$
as $| Z \rangle = | \bar\phi_{34} \rangle = \mathfrak{c}^{2 \dagger} \mathfrak{d}^{2\dagger}| 0
\rangle$.} $| Z \rangle$ identified with particle content of the $sl(2|2)$ subsector as follows,
\be
\mathfrak{a}^\dagger \mathfrak{b}^\dagger | Z \rangle = | \mathcal{D}^+ Z \rangle
\, , \qquad
\mathfrak{c}^{1\dagger} \mathfrak{c}_2 | Z \rangle = | X \rangle
\, , \qquad
\mathfrak{a}^\dagger \mathfrak{c}_2 | Z \rangle = | \bar\chi \rangle
\, , \qquad
\mathfrak{b}^\dagger \mathfrak{c}^{1 \dagger} | Z \rangle = | \psi \rangle
\, .
\ee
The one-site ground state and one-site elementary excitations have the following nonvanishing chirality
and isospin
\be
\mathcal{B} | \bar{\chi} \rangle = - \ft14 | \bar{\chi} \rangle
\, , \qquad
\mathcal{B} | \psi \rangle = \ft14 | \psi \rangle
\, , \qquad
\mathcal{T}^0 | X \rangle = \ft12 | X \rangle
\, , \qquad
\mathcal{T}^0 | Z \rangle = - \ft12 | Z \rangle
\, .
\ee

The ground state of the second level in the nested Bethe Ansatz is chosen in terms of noncompact primary
excitations $\mathcal{D}^+ Z (0)$ on each site of the chain. The third vacuum state can be chosen in term
of fermions, either $\bar{\chi}$ or $\psi$. Thus the excitations on the spin chain sites are built in
terms of raising operators corresponding to simple roots acting on nodes of the Kac-Dynkin diagram such that
a general state reads schematically
\ba
\label{ExcitationState}
| \mathcal{O}_{\bit{\scriptstyle \omega}} \rangle
=
( e^+_3 )^{n_3}
( e^+_1 )^{n_1}
( e^+_2 )^{n_2}
| \Omega \rangle_L
\!\!\!&=&\!\!\!
( \mathcal{V}_2^- )^{n_3}
( - \bar{\mathcal{V}}^{1, -} )^{n_1}
( \mathcal{L}^+ )^{n_2}
| \Omega \rangle_L
\\
&\sim&\!\!\!
( \mathfrak{c}_2 )^{n_3}
( \mathfrak{c}^{1 \dagger} )^{n_1}
( \mathfrak{a}^\dagger )^{n_2 - n_1}
( \mathfrak{b}^\dagger )^{n_2 - n_3}
| \Omega \rangle_L
\, . \nonumber
\ea
This formula requires clarifications. First, since the fermionic generators are nilpotent they all
have to act on different spin chain sites, thus their number cannot exceed the number of sites, i.e.,
$0 \leq n_{1,3} \leq L$. For instance, for $n_3 = 0$ and $n_1 = L$ and $n_2 \geq n_1$, the resulting
state corresponds to the $L-$fermion operator $(\mathcal{D}^+)^{n_2 - n_1} \psi^L (0)$.  Second,
since both $e^+_1$ and $e^+_3$ annihilate the vacuum state $| \Omega \rangle_L$, this gives a natural
restriction on the number of excitations acting on the same sites, i.e., $n_1 \leq n_2 \geq n_3$.
Third, the number of noncompact excitations $n_2$ is unrestricted from above. Now, the number-of-excitation
operators
\be
\mathcal{N}_{\mathfrak{a}} = \mathfrak{a}^\dagger \mathfrak{a}
\, , \qquad
\mathcal{N}_{\mathfrak{b}} = \mathfrak{b}^\dagger \mathfrak{b}
\, , \qquad
\mathcal{N}_{\mathfrak{c_1}} = \mathfrak{c}^{1 \dagger} \mathfrak{c}_1
\, , \qquad
\mathcal{N}_{\mathfrak{c_2}} = \mathfrak{c}_2 \mathfrak{c}^{2 \dagger}
\, ,
\ee
can be related to generators of the Cartan subalgebra of $psl(2|2)$ and $u_B (1)$ automorphism
\be
\mathcal{N}_{\mathfrak{a}} = \mathcal{L}^0 - 2 \mathcal{B} - \ft12 L
\, , \quad
\mathcal{N}_{\mathfrak{b}} = \mathcal{L}^0 + 2 \mathcal{B} - \ft12 L
\, , \quad
\mathcal{N}_{\mathfrak{c}_1} = \mathcal{T}^0 + 2 \mathcal{B} + \ft12 L
\, , \quad
\mathcal{N}_{\mathfrak{c}_2} = \mathcal{T}^0 - 2 \mathcal{B} + \ft12 L
\, , \quad
\ee
and yield the following relations between the excitation numbers $n_p$ and eigenvalues $\ell$, $t$ and $b$
of $\mathcal{L}^0$, $\mathcal{B}$ and $\mathcal{T}^0$, respectively, for the state \re{ExcitationState}
\be
n_1
= n_{\mathfrak{c}_1}
= t + 2 b + \ft12 L
\, , \quad
n_2
=
n_{\mathfrak{a}} + n_{\mathfrak{c}_1}
=
n_{\mathfrak{b}} + n_{\mathfrak{c}_2}
=
t + \ell
\, , \quad
n_3
= n_{\mathfrak{c}_2}
= t - 2 b + \ft12 L
\, .
\ee

\section{Transfer matrices for low-dimensional representations}
\label{TransferAppendix}

We give here explicit expressions for transfer matrices with low-dimensional auxiliary space.
The transfer matrix conjugate to Eq.\ \re{BaxterQ2} reads
\ba
\bar{t}_{[1]} (x)
\!\!\!&=&\!\!\!
(x^+)^L
\frac{\widehat{Q}^{(3)} \left( u - \ft{i}{2}\right)}{\widehat{Q}^{(3)} \left( u + \ft{i}{2}\right)}
\left(
{\rm e}^{\Delta_+ (x^+)} \frac{Q^{(2)} (u + i)}{Q^{(2)} (u)}
-
{\rm e}^{\frac{1}{2} \Delta_- (x^+) + \frac{1}{2} \Delta_+ (x^+)}
\right)
\nonumber\\
&+&\!\!\!
(x^-)^L
\frac{\widehat{Q}^{(1)} \left( u + \ft{i}{2}\right)}{\widehat{Q}^{(1)} \left( u - \ft{i}{2}\right)}
\left(
{\rm e}^{\Delta_- (x^-)} \frac{Q^{(2)} (u - i)}{Q^{(2)} (u)}
-
{\rm e}^{\frac{1}{2} \Delta_- (x^-) + \frac{1}{2} \Delta_+ (x^-)}
\right)
\, .
\ea
The generating function \re{GFantiAllOrder} yields the eigenvalues of antisymmetric transfer
matrix $t_{[2]}$
\ba
&&
{\rm e}^{- \ft12 \Delta_- (x) - \ft12 \Delta_+ (x)} t_{[2]} (x)
\\
&&
\qquad
= \left( x \, x^{[-2]} \right)^L
\frac{Q^{(2)} \left( u - \ft{i}{2}\right)}{Q^{(2)} \left( u + \ft{i}{2}\right)}
\frac{\widehat{Q}^{(3)} (u + i)}{\widehat{Q}^{(3)} (u - i)}
\left(
{\rm e}^{\frac{1}{2} \Delta_+ \left(x^{[- 2]}\right) + \frac{1}{2} \Delta_- \left(x^{[- 2]}\right)}
-
{\rm e}^{\Delta_- \left(x^{[-2]}\right)} \frac{Q^{(2)} \left(u - \ft{3 i}{2}\right)}{Q^{(2)} \left(u - \ft{i}{2}\right)}
\right)
\nonumber\\
&&\qquad
+
\left( x \, x^{[+2]} \right)^L
\frac{Q^{(2)} \left( u + \ft{i}{2}\right)}{Q^{(2)} \left( u - \ft{i}{2}\right)}
\frac{\widehat{Q}^{(1)} (u - i)}{\widehat{Q}^{(1)} (u + i)}
\left(
{\rm e}^{\frac{1}{2} \Delta_+ \left(x^{[+ 2]}\right) + \frac{1}{2} \Delta_- \left(x^{[+ 2]}\right)}
-
{\rm e}^{\Delta_+ \left(x^{[+ 2]}\right)} \frac{Q^{(2)} \left(u + \ft{3 i}{2}\right)}{Q^{(2)} \left(u + \ft{i}{2}\right)}
\right)
\nonumber\\
&&\qquad
-
x^{2 L}
\frac{\widehat{Q}^{(3)} (u + i)}{\widehat{Q}^{(3)} (u)}
\frac{\widehat{Q}^{(1)} (u - i)}{\widehat{Q}^{(1)} (u)}
\frac{
\Big(
{\rm e}^{\ft12 \Delta_+ (x)} Q^{(2)} \left(u + \ft{i}{2}\right)
-
{\rm e}^{\ft12 \Delta_- (x)} Q^{(2)} \left(u - \ft{i}{2}\right)
\Big)^2
}{Q^{(2)} \left(u + \ft{i}{2}\right) Q^{(2)} \left(u - \ft{i}{2}\right)}
\, , \nonumber
\ea
and its conjugate $\bar{t}_{[2]}$
\ba
&&
{\rm e}^{- \ft12 \Delta_- (x) - \ft12 \Delta_+ (x)} \bar{t}_{[2]} (x)
\\
&&
\qquad
= \left( x \, x^{[+2]} \right)^L
\frac{Q^{(2)} \left( u + \ft{i}{2}\right)}{Q^{(2)} \left( u - \ft{i}{2}\right)}
\frac{\widehat{Q}^{(3)} (u - i)}{\widehat{Q}^{(3)} (u + i)}
\left(
{\rm e}^{\frac{1}{2} \Delta_- \left(x^{[+ 2]}\right) + \frac{1}{2} \Delta_+ \left(x^{[+ 2]}\right)}
-
{\rm e}^{\Delta_+ \left(x^{[+ 2]}\right)} \frac{Q^{(2)} \left(u + \ft{3 i}{2}\right)}{Q^{(2)} \left(u + \ft{i}{2}\right)}
\right)
\nonumber\\
&&\qquad
+
\left( x \, x^{[- 2]} \right)^L
\frac{Q^{(2)} \left( u - \ft{i}{2}\right)}{Q^{(2)} \left( u + \ft{i}{2}\right)}
\frac{\widehat{Q}^{(1)} (u + i)}{\widehat{Q}^{(1)} (u - i)}
\left(
{\rm e}^{\frac{1}{2} \Delta_- \left(x^{[- 2]}\right) + \frac{1}{2} \Delta_+ \left(x^{[- 2]}\right)}
-
{\rm e}^{\Delta_- \left(x^{[- 2]}\right)} \frac{Q^{(2)} \left(u - \ft{3 i}{2}\right)}{Q^{(2)} \left(u - \ft{i}{2}\right)}
\right)
\nonumber\\
&&\qquad
-
x^{2 L}
\frac{\widehat{Q}^{(3)} (u - i)}{\widehat{Q}^{(3)} (u)}
\frac{\widehat{Q}^{(1)} (u + i)}{\widehat{Q}^{(1)} (u)}
\frac{
\Big(
{\rm e}^{\ft12 \Delta_- (x)} Q^{(2)} \left(u - \ft{i}{2}\right)
-
{\rm e}^{\ft12 \Delta_+ (x)} Q^{(2)} \left(u + \ft{i}{2}\right)
\Big)^2
}{Q^{(2)} \left(u - \ft{i}{2}\right) Q^{(2)} \left(u + \ft{i}{2}\right)}
\, . \nonumber
\ea
Similarly, one finds from Eq.\ \re{GFsymmAllOrder} for symmetric $t^{\{2\}}$
\ba
t^{\{2\}} (x)
\!\!\!&=&\!\!\!
\left( x^{[-2]} x^{[+2]} \right)^L
\left(
{\rm e}^{\Delta_- \left(x^{[-2]}\right)} \frac{Q^{(2)} \left(u - \ft{3 i}{2}\right)}{Q^{(2)} \left(u - \ft{i}{2}\right)}
-
{\rm e}^{\frac{1}{2} \Delta_+ \left(x^{[- 2]}\right) + \frac{1}{2} \Delta_- \left(x^{[- 2]}\right)}
\right)
\\
&&\qquad\qquad
\times
\left(
{\rm e}^{\Delta_+ \left(x^{[+2]}\right)} \frac{Q^{(2)} \left(u + \ft{3 i}{2}\right)}{Q^{(2)} \left(u + \ft{i}{2}\right)}
-
{\rm e}^{\frac{1}{2} \Delta_+ \left(x^{[+2]}\right) + \frac{1}{2} \Delta_- \left(x^{[+2]}\right)}
\right)
\frac{\widehat{Q}^{(3)} (u)}{\widehat{Q}^{(3)} (u - i)}
\frac{\widehat{Q}^{(1)} (u)}{\widehat{Q}^{(1)} (u + i)}
\nonumber\\
&&\!\!\!\!\!\!\!\!\!\!\!\!\!\!\!\!\!\!\!\!\!
+
\left( x \, x^{[-2]} \right)^L
{\rm e}^{\Delta_- (x)}
\left(
{\rm e}^{\Delta_- \left(x^{[-2]}\right)} \frac{Q^{(2)} \left(u - \ft{3 i}{2}\right)}{Q^{(2)} \left(u - \ft{i}{2}\right)}
-
{\rm e}^{\frac{1}{2} \Delta_+ \left(x^{[- 2]}\right) + \frac{1}{2} \Delta_- \left(x^{[- 2]}\right)}
\right)
\frac{Q^{(2)} \left( u - \ft{i}{2}\right)}{Q^{(2)} \left( u + \ft{i}{2}\right)}
\frac{\widehat{Q}^{(3)} (u + i)}{\widehat{Q}^{(3)} (u - i)}
\nonumber\\
&&\!\!\!\!\!\!\!\!\!\!\!\!\!\!\!\!\!\!\!\!\!
+
\left( x \, x^{[+2]} \right)^L
{\rm e}^{\Delta_+ (x)}
\left(
{\rm e}^{\Delta_+ \left(x^{[+2]}\right)} \frac{Q^{(2)} \left(u + \ft{3 i}{2}\right)}{Q^{(2)} \left(u + \ft{i}{2}\right)}
-
{\rm e}^{\frac{1}{2} \Delta_+ \left(x^{[+2]}\right) + \frac{1}{2} \Delta_- \left(x^{[+2]}\right)}
\right)
\frac{Q^{(2)} \left( u + \ft{i}{2}\right)}{Q^{(2)} \left( u - \ft{i}{2}\right)}
\frac{\widehat{Q}^{(1)} (u - i)}{\widehat{Q}^{(1)} (u + i)}
, \nonumber
\ea
and $\bar{t}^{\{2\}}$
\ba
\bar{t}^{\{2\}} (x)
\!\!\!&=&\!\!\!
\left( x^{[+2]} x^{[-2]} \right)^L
\left(
{\rm e}^{\Delta_+ \left(x^{[+2]}\right)} \frac{Q^{(2)} \left(u + \ft{3 i}{2}\right)}{Q^{(2)} \left(u + \ft{i}{2}\right)}
-
{\rm e}^{\frac{1}{2} \Delta_- \left(x^{[+2]}\right) + \frac{1}{2} \Delta_+ \left(x^{[+2]}\right)}
\right)
\\
&&\qquad\qquad
\times
\left(
{\rm e}^{\Delta_- \left(x^{[-2]}\right)} \frac{Q^{(2)} \left(u - \ft{3 i}{2}\right)}{Q^{(2)} \left(u - \ft{i}{2}\right)}
-
{\rm e}^{\frac{1}{2} \Delta_- \left(x^{[-2]}\right) + \frac{1}{2} \Delta_+ \left(x^{[-2]}\right)}
\right)
\frac{\widehat{Q}^{(3)} (u)}{\widehat{Q}^{(3)} (u + i)}
\frac{\widehat{Q}^{(1)} (u)}{\widehat{Q}^{(1)} (u - i)}
\nonumber\\
&&\!\!\!\!\!\!\!\!\!\!\!\!\!\!\!\!\!\!\!\!\!
+ \left( x \, x^{[+2]} \right)^L
{\rm e}^{\Delta_+ (x)}
\left(
{\rm e}^{\Delta_+ \left(x^{[+2]}\right)} \frac{Q^{(2)} \left(u + \ft{3 i}{2}\right)}{Q^{(2)} \left(u + \ft{i}{2}\right)}
-
{\rm e}^{\frac{1}{2} \Delta_- \left(x^{[+2]}\right) - \frac{1}{2} \Delta_+ \left(x^{[+2]}\right)}
\right)
\frac{Q^{(2)} \left( u + \ft{i}{2}\right)}{Q^{(2)} \left( u - \ft{i}{2}\right)}
\frac{\widehat{Q}^{(3)} (u - i)}{\widehat{Q}^{(3)} (u + i)}
\nonumber\\
&&\!\!\!\!\!\!\!\!\!\!\!\!\!\!\!\!\!\!\!\!\!
+
\left( x \, x^{[-2]} \right)^L
{\rm e}^{\Delta_- (x)}
\left(
{\rm e}^{\Delta_- \left(x^{[-2]}\right)} \frac{Q^{(2)} \left(u - \ft{3 i}{2}\right)}{Q^{(2)} \left(u - \ft{i}{2}\right)}
-
{\rm e}^{\frac{1}{2} \Delta_- \left(x^{[-2]}\right) + \frac{1}{2} \Delta_+ \left(x^{[-2]}\right)}
\right)
\frac{Q^{(2)} \left( u - \ft{i}{2}\right)}{Q^{(2)} \left( u + \ft{i}{2}\right)}
\frac{\widehat{Q}^{(1)} (u + i)}{\widehat{Q}^{(1)} (u - i)}
.
\nonumber
\ea



\end{document}